\documentclass[journal,onecolumn]{IEEEtran}
   \usepackage{graphicx}
\usepackage[cmex10]{amsmath}
\usepackage{multicol}

\usepackage[tight,footnotesize]{subfigure}
\begin{document}
 
\title{\LARGE \bf \copyright 2013 IEEE. Personal use of this material is permitted. Permission from IEEE must be obtained for all other uses, in any current or future media, including reprinting/republishing this material for advertising or promotional purposes, creating new collective works, for resale or distribution to servers or lists, or reuse of any copyrighted component of this work in other works. The complete paper can be found at: \url{http://ieeexplore.ieee.org/xpl/articleDetails.jsp?tp=&arnumber=6607217} and Digital Object Identifier :10.1109/TAP.2013.2283270 \\
\vspace{30em}
A Link Loss Model for the On-body Propagation Channel for Binaural Hearing Aids}
%
%
%

\author{Rohit Chandra and Anders J Johansson%
\thanks{Rohit Chandra and Anders J Johansson are with the Department
of Electrical and Information Technology, Lund University, Lund,
Sweden, e-mail: rohit.chandra@eit.lth.se}
}

\maketitle

\begin{abstract}
Binaural hearing aids communicate with each other through a wireless link for synchronization. A propagation model is needed to estimate the ear-to-ear link loss for such binaural hearing aids. The link loss is a critical parameter in a link budget to decide the sensitivity of the transceiver. In this paper, we have presented a model for the deterministic component of the ear-to-ear link loss. The model takes into account the dominant paths having most of the power of the creeping wave from the transceiver in one ear to the transceiver in other ear and the effect of the protruding part of the outer ear called pinna. Simulations are done to validate the model using in-the-ear (ITE) placement of antennas at 2.45 GHz on two heterogeneous phantoms of different age-group and body size. The model agrees with the simulations. The ear-to-ear link loss between the antennas for the binaural hearing aids in the homogeneous SAM phantom is compared with a heterogeneous phantom. It is found that the absence of the pinna and the lossless shell in the SAM phantom underestimate the link loss. This is verified by the measurements on a phantom where we have included the pinnas fabricated by 3D-printing. 
\end{abstract}



%
\IEEEpeerreviewmaketitle

\section{Introduction}
%
%
%
%
\IEEEPARstart{W}{earable} medical devices have revolutionized the field of medical sciences and have improved the quality of life of both diseased and healthy persons. They can be used for vital signal monitoring of patients as in Wireless Body Area Networks (WBAN) and can alert doctors about the critical condition of their patients. These devices can be used for rehabilitation of patients with disability, and also by healthy persons without any disease or disability e.g. by sportsmen and athletes. The main advantage of the wearable devices is that they are non-invasive and have limited risk of infection \cite{pf},\cite{dittmar}. 

The wearable medical devices placed at different positions on the body may communicate with each other for data exchange. In such scenarios, on-body propagation models become critical for proper estimation of the link budget. Various statistical and analytical on-body propagation models have been discussed in \cite{cotton}-\cite{conway}. In \cite{cotton}, a time domain analysis and modeling of the on-body propagation characteristics was presented. The measurements were done in different scenarios like anechoic chamber, open office area, hallway, and outdoor environment and autocorrelation and cross-correlation statistics were presented at $2.45$ GHz. In \cite{fort}, a simplified physical body area propagation model was derived using Maxwell's equations. A channel model for wireless communication around the human body was developed in \cite{ryk}. The creeping wave is the phenomenon for the communication between the transceivers located on the opposite side of the body. An analytical propagation model of BAN channels based on the creeping wave theory was presented in \cite{alves} by re-using the theory of the wave propagation over the inhomogeneous earth's surface \cite{wait}.

Hearing aids are one such wearable medical device which is used for rehabilitation of hearing impaired persons. Binaural hearing aids are a system where there is a hearing aid in both the ears of the user. They communicate with each other for synchronization data \cite{chandra1} through the ear-to-ear on-body propagation channel. The ear-to-ear propagation channel has been discussed in \cite{zaso}-\cite{kvist3}. In \cite{zaso} it was shown that diffraction is the main propagation mechanism around the head for the link between the ears. This diffraction mechanism is the same as the creeping wave phenomenon. It was shown in our previous work \cite{chandra1} that in-the-ear (ITE) placement has less attenuation than in-the-canal (ITC) placement of the binaural hearing aids as the communication between the ITE antennas was through the creeping waves. In \cite{kvist1}, the effect of the head size on the ear-to-ear radio propagation channel was examined. It was shown by the simulations that the variations in the head size may result in up to $10$ dB variation in the link loss. In \cite{bour}, authors have shown by measurements with different antennas on the SAM head that small variations in the position of the antenna do not affect the ear-to-ear link loss whereas the operating frequencies have a larger impact on the link loss. Their result showed that the link loss increases with increase in the frequency. The importance of the correct orientation of the antenna to decrease the link loss was also shown. Homogeneous models were used for the simulations or for the measurements. The protruding part of the outer ear called pinna and the lossy skin are absent in the SAM head/homogeneous head. However, in \cite{chandra2} we have shown the significant effect of the pinna and the lossy skin on the ear-to-ear link loss. The propagation model for the communication between the ITE binaural hearing aids is required for proper estimation of the ear-to-ear link loss. The model for the propagation around the head presented in \cite{alves}, can be used for this purpose. Since the model was developed for a simplified model of the head, treating it as a cylinder, there was a good agreement between the simulations and the measurements. However, this model does not take into account the losses due to the pinnas. 

In this paper, we have presented an analytical model which includes the losses of the pinnas. The cross-section of the head is modeled as a more realistic elliptical shape rather than a circular shape. It is shown that the two paths of the creeping waves are sufficient to describe the link loss. Simulations are done in SEMCAD-X \cite{semcad} which uses the FDTD method. A low-profile and compact ITE antenna \cite{chandra2} on the heterogeneous phantoms is used to verify the propagation model. Measurements are done to verify the effect of the pinna on the ear-to-ear link loss. Both the simulations and the measurements are done in $2.45$ GHz ISM band. 

The paper is organized as follows. In Section \ref{sec:num_phantom}, the numerical phantoms used for the simulations are described. Section \ref{sec:model} present the analytical model. Other factors apart from the pinna which may affect the ear-to-ear link loss are discussed in Section \ref{sec:factors}. In Section \ref{sec:meas}, the measurement process and results are presented. Conclusions are presented in Section \ref{sec:conclusion}.

\section{Numerical Phantoms} \label{sec:num_phantom}

\begin{figure}
  \begin{center}
    \subfigure[]{\label{SAM_Phantom_dim}\includegraphics[scale=0.13]{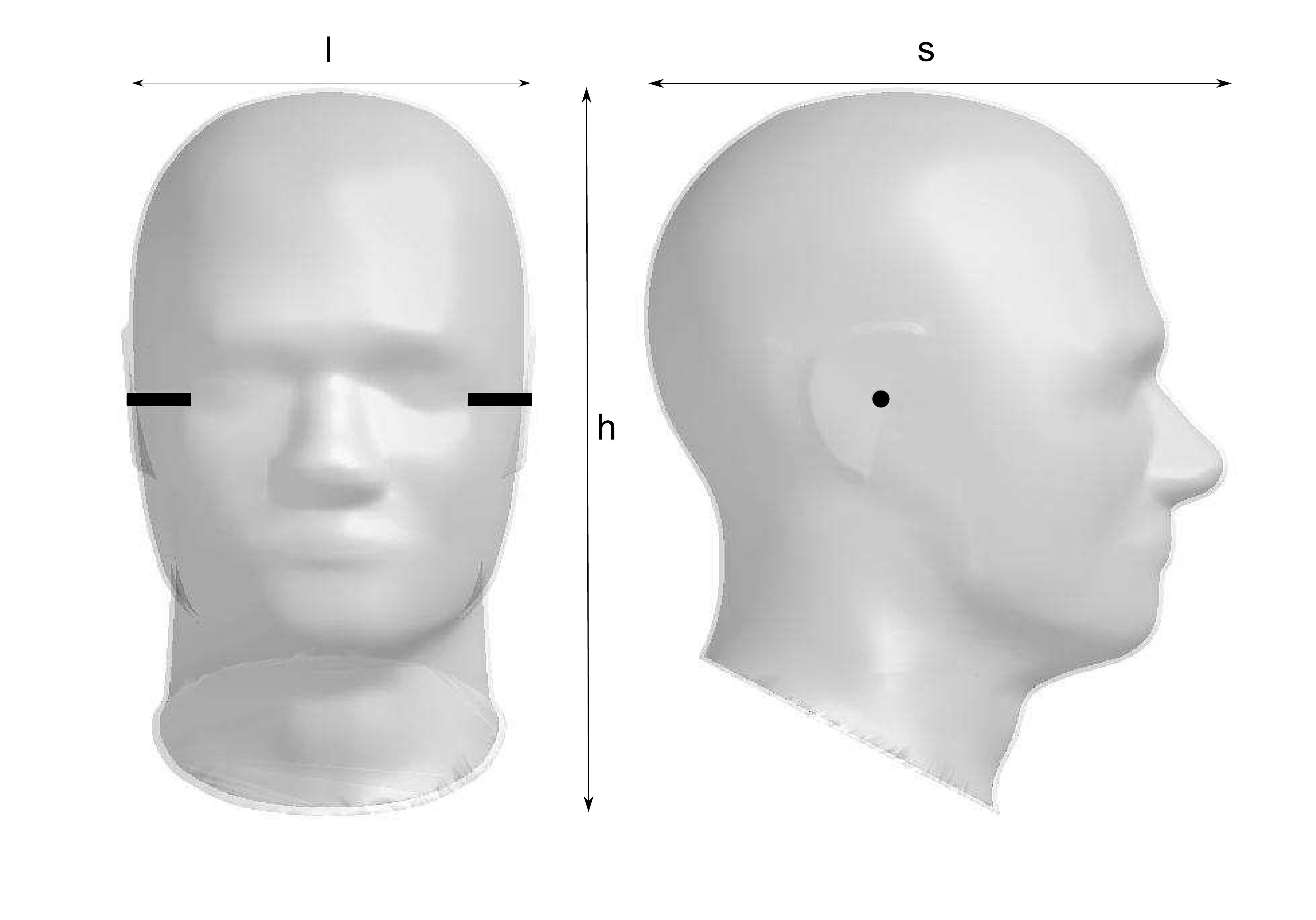}}\\
    \subfigure[]{\label{family_heads_dim}\includegraphics[scale=0.28]{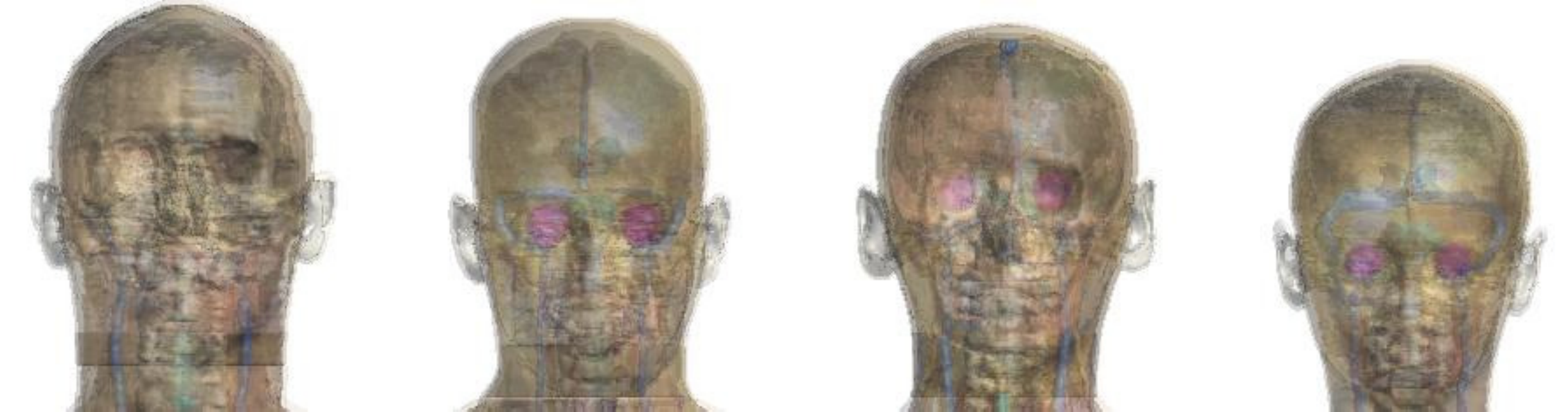}}%
  \end{center}
  \caption{Phantoms (a) SAM Head (b) Heterogeneous Phantoms (only the truncated head part is shown here). From L to R: Duke, Ella, Louis, Billie}
  \label{family_heads_total}
\end{figure}

Four numerical heterogeneous phantoms of different age group and gender and one homogeneous phantom are used for the simulations. The modified SAM head phantom with the ear canals~\cite{chandra1}, shown in Fig.~\ref{SAM_Phantom_dim}, is used as a homogeneous phantom. The heterogeneous numerical phantoms used in this paper are provided by ITIS foundation~\cite{itis} developed for the Virtual Family and Classroom project~\cite{vfp}. They are whole body anatomical models consisting of more than $80$ tissues with different electrical properties. The models are that of a $34$ year old male called Duke, a $26$ year old female called Ella, a $15$ year old boy, Louis and a $11$ year old girl, Billie. They are shown in Fig.~\ref{family_heads_dim}. Since we are interested in the ear-to-ear link loss, only the truncated head with the neck of the phantoms are used for the simulations. This also has an advantage in terms of faster simulation. However, we have examined and presented the effect of the shoulders on the link loss. Table~\ref{tissue_parameter} presents the electrical properties of the tissues present in the head and the neck of the phantoms at $2.45$ GHz. Tissue is the name of the tissue of the phantom of Virtual Family Project and Gabriel list gives the mapping of the phantom tissue to the Gabriel list obtained from~\cite{ifac} for the electrical properties. The approximate head dimensions of the phantoms are shown in Table~\ref{head_dimension} where the dimensions $l, h, m$ are shown in Fig.~\ref{family_heads_total}.

\begin{table}[ht]\scriptsize
\caption{Tissue Parameters}
\label{tissue_parameter}
\begin{center}
\begin{tabular}{|c||c||c||c|}
\hline
 \bf {Tissue} & \bf{Gabriel List} & \bf{$\varepsilon_r$} & \bf{$\sigma_e$ (S/m)}\\
\hline
Air Internal & Air & 1 & 0\\
 \hline
Artery & Blood & 58.26 & 2.54\\
\hline
Blood Vessel & Blood & 58.26 & 2.54 \\
\hline
Bone & Bone Cortical & 11.38 & 0.39 \\
\hline
Brain Grey Matter & Brain Grey Matter & 48.91 & 1.80 \\
\hline
Brain White Matter & Brain White Matter & 36.16 & 1.21 \\
\hline
Cartilage & Cartilage & 38.77 & 1.75 \\
\hline
Cerebellum & Cerebellum & 44.80 & 2.10 \\
\hline
Cerebrospinal fluid & Celebro Spinal Fluid & 66.24 & 3.45 \\
\hline
Commissura anterior & Brain White Matter & 36.16 & 1.21\\
\hline
Commissura posterior & Brain White Matter & 36.16 & 1.21\\
\hline
Connective tissue & Fat(mean) & 10.82 & 0.26\\
\hline
Cornea & Cornea & 51.61 & 2.29\\
\hline
Ear Cartilage & Cartilage & 38.77 & 1.75\\
\hline
Ear Skin & Skin(Dry) & 38.00 & 1.46\\
\hline
Eye Lens & Lens Cortex & 44.62 & 1.50\\
\hline
Eye Sclera & Eye Tissue (Sclera) & 52.62 & 2.03\\
\hline
Eye vitreous humor & Vitreous Humor & 68.20 & 2.47\\
\hline
Fat & Fat & 5.28 & 0.10\\
\hline
Hippocampus & Brain Grey Matter & 48.91 & 1.80\\
\hline
Hypophysis & Gland & 57.20 & 1.96\\
\hline
Hypothalamus & Gland & 57.20 & 1.96\\
\hline
Intervertebral disc & Cartilage & 38.77 & 1.75\\
\hline
Larynx & Cartilage & 38.77 & 1.75\\
\hline
Mandible & Bone Cortical & 11.38 & 0.39\\
\hline
Marrow Red & Bone Marrow(infiltrated) & 10.30 & 0.45\\
\hline
Medulla Oblongata & Brain(average) & 42.53 & 1.51\\
\hline
Midbrain & Brain(average) & 42.53 & 1.51\\
\hline
Mucosa & Mucous Membrane & 42.85 & 1.59\\
\hline
Muscle & Muscle(Transverse Fiber) & 38.00 & 1.46\\
\hline
Nerve & Nerve & 30.14 & 1.08\\
\hline
Pharynx & Air & 1 & 0\\
\hline
SAT & Fat & 5.28 & 0.10\\
\hline
Skin & Skin(Dry) & 38.00 & 1.46\\
\hline
Skull & Bone Cortical & 11.38 & 0.39\\
\hline
Spinal Cord & Spinal Chord & 30.14 & 1.08\\
\hline
Teeth & Tooth & 11.38 & 0.39\\
\hline
Tendon Ligament & Tendon & 43.12 & 1.68\\
\hline
Thalamus & Brain Grey Matter & 48.91 & 1.80\\
\hline
Thymus & Thymus & 57.20 & 1.96\\
\hline
Thyroid gland & Gland & 57.20 & 1.96\\
\hline
Tongue & Tongue & 52.62 & 1.80\\
\hline
Vein & Blood & 58.26 & 2.54\\
\hline
Vertebrae & Bone Cortical & 11.38 & 0.39\\
\hline
\end{tabular}
\end{center}
\end{table}

\begin{table}[ht]\scriptsize
\caption{Dimensions of the phantom's head}
\label{head_dimension}
\begin{center}
\begin{tabular}{|c||c||c||c||c||c|}
\hline 
Phantom				&		Age(year)	 &	 Gender		&		l(mm)		&		h(mm)		&		s(mm)		\\
\hline
SAM						&			--			 &    Male		&		164			&		312			&		234			\\
\hline
Duke					&			34			 &	  Male		&		155			&		251 		&		238			\\
\hline
Ella					&			26			 &    Female	&		143			&		239		  &		209			\\
\hline
Louis					&			15			 &		Male		&		152			&		241		  &		214			\\
\hline
Billie				&			11			 &		Female  &		137			&		233		  &		188			\\			
\hline
\end{tabular}
\end{center}
\scriptsize{For l, h and m see Fig.~\ref{family_heads_total}}
\end{table} 

\section{Link Loss Model}\label{sec:model}
It is well established that the creeping wave is the dominant phenomenon for the communication around the curved part of the body. In \cite{alves} and \cite{fort1}, it was shown that the clockwise and the anti-clockwise creeping wave interfere with each other. The phantom in these investigations was symmetrical. However, the realistic phantoms are anatomically asymmetric and hence for the ear-to-ear channel there will be several paths such as over top of the head, over back of the head, over the front part of the head, etc. Investigations are done through simulations to determine the dominant paths i.e. the ones having most of the power of the creeping waves.  

\subsection{Dominant Paths of the Creeping Waves}
\begin{figure}[thpb]
 			\begin{center}
    			\subfigure[]{\label{case1}\includegraphics[scale=0.075]{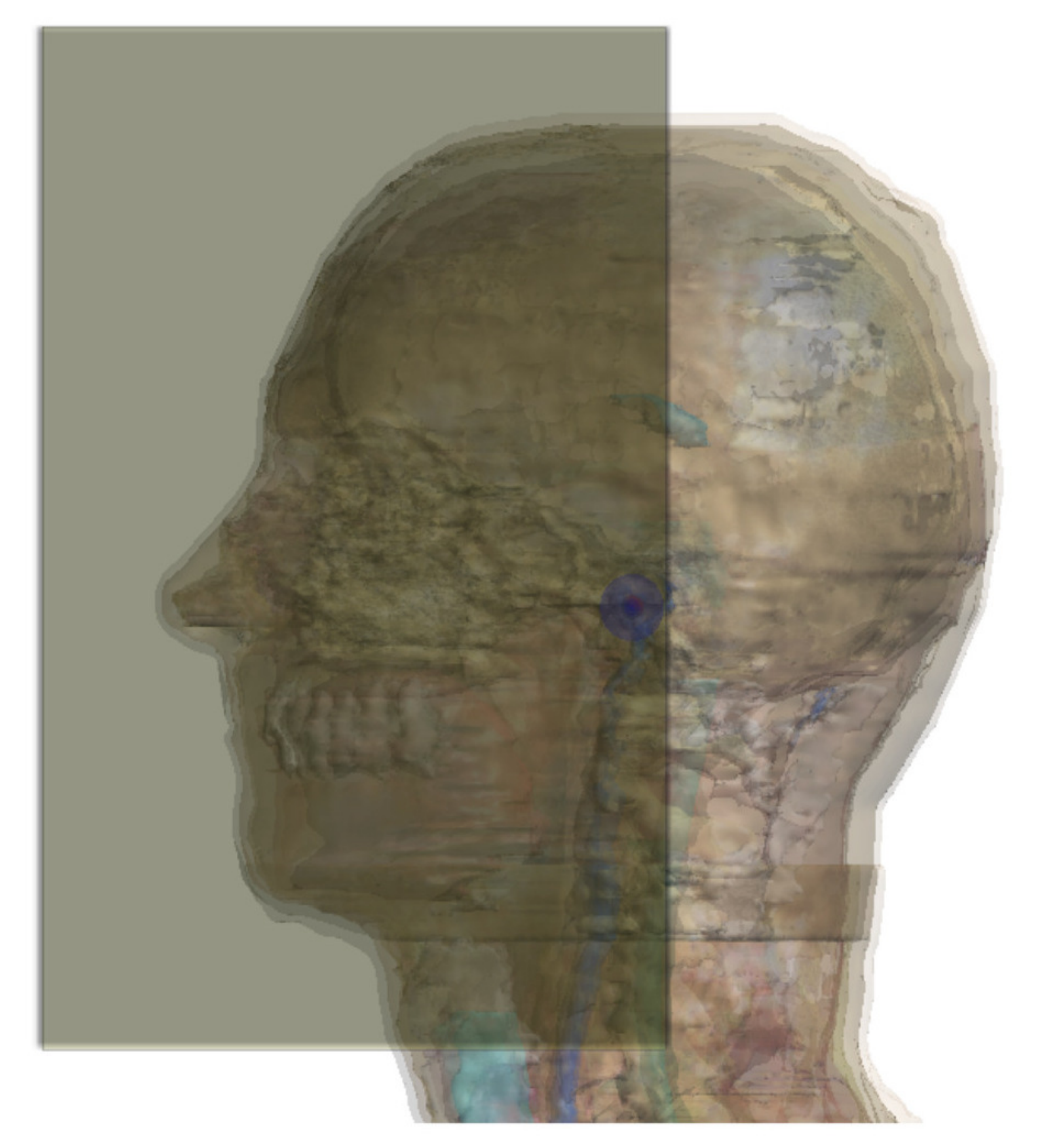}}
    			\subfigure[]{\label{case2}\includegraphics[scale=0.075]{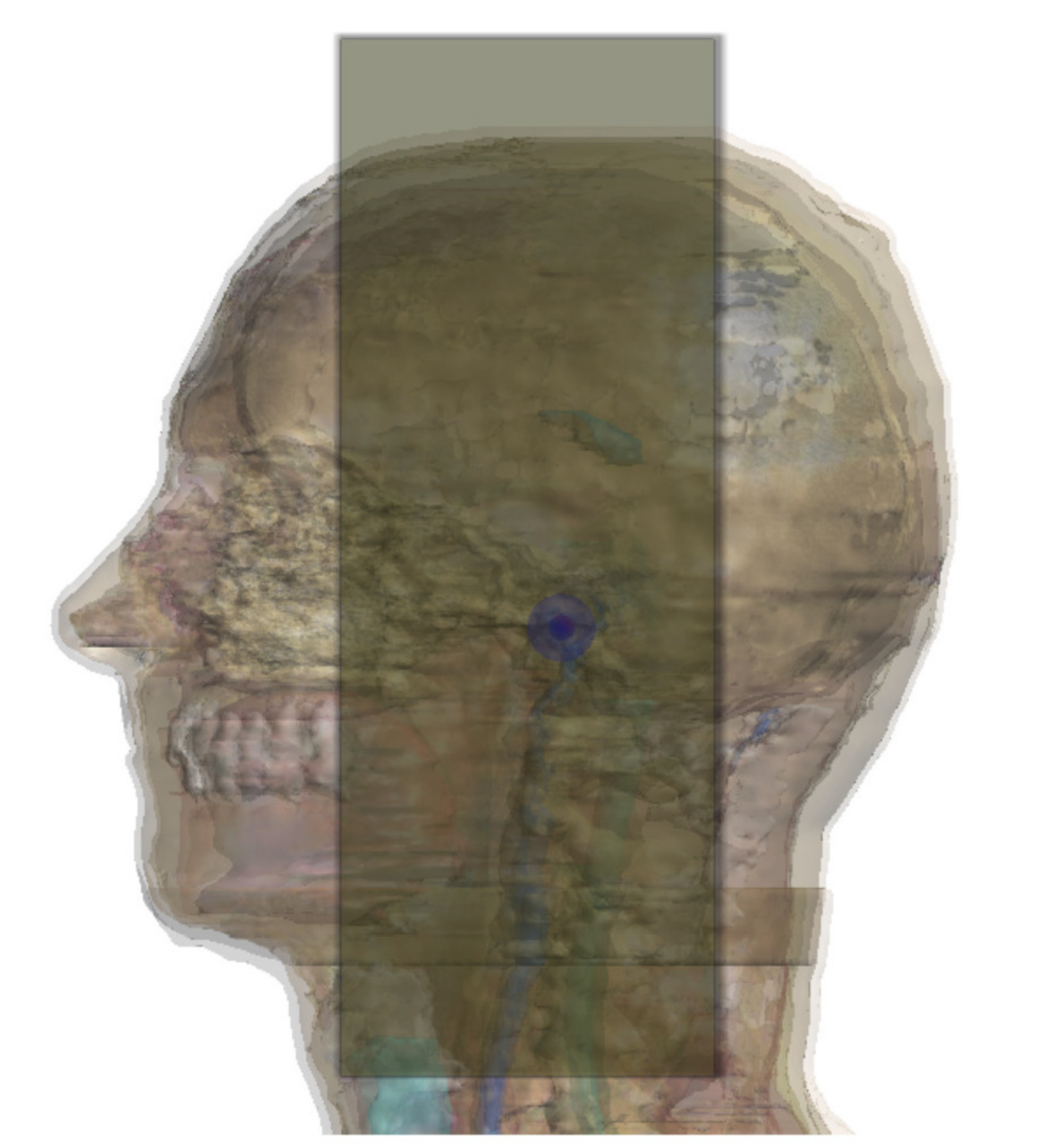}}
    			\subfigure[]{\label{case3}\includegraphics[scale=0.075]{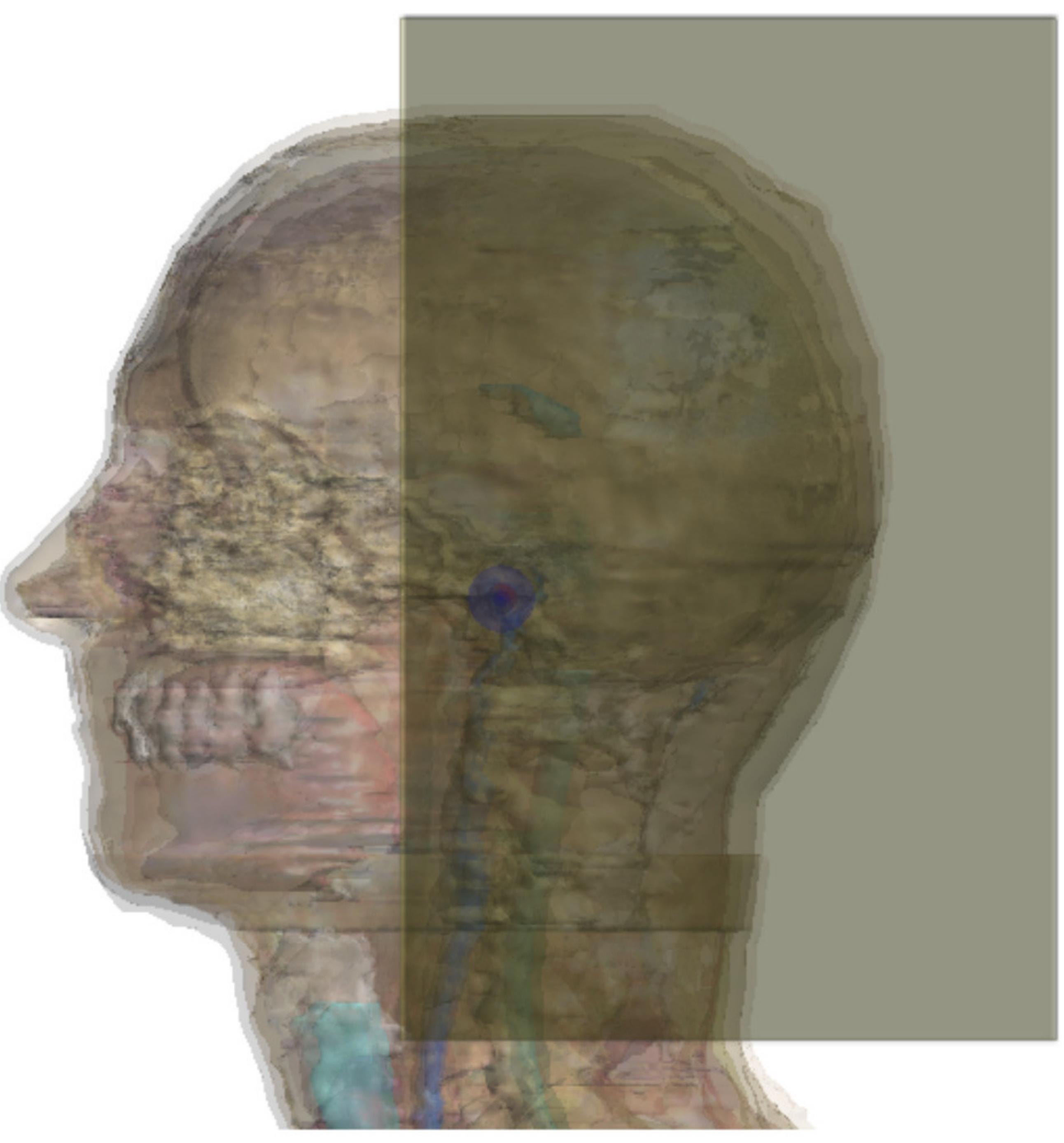}}\\
    			\subfigure[]{\label{case4}\includegraphics[scale=0.075]{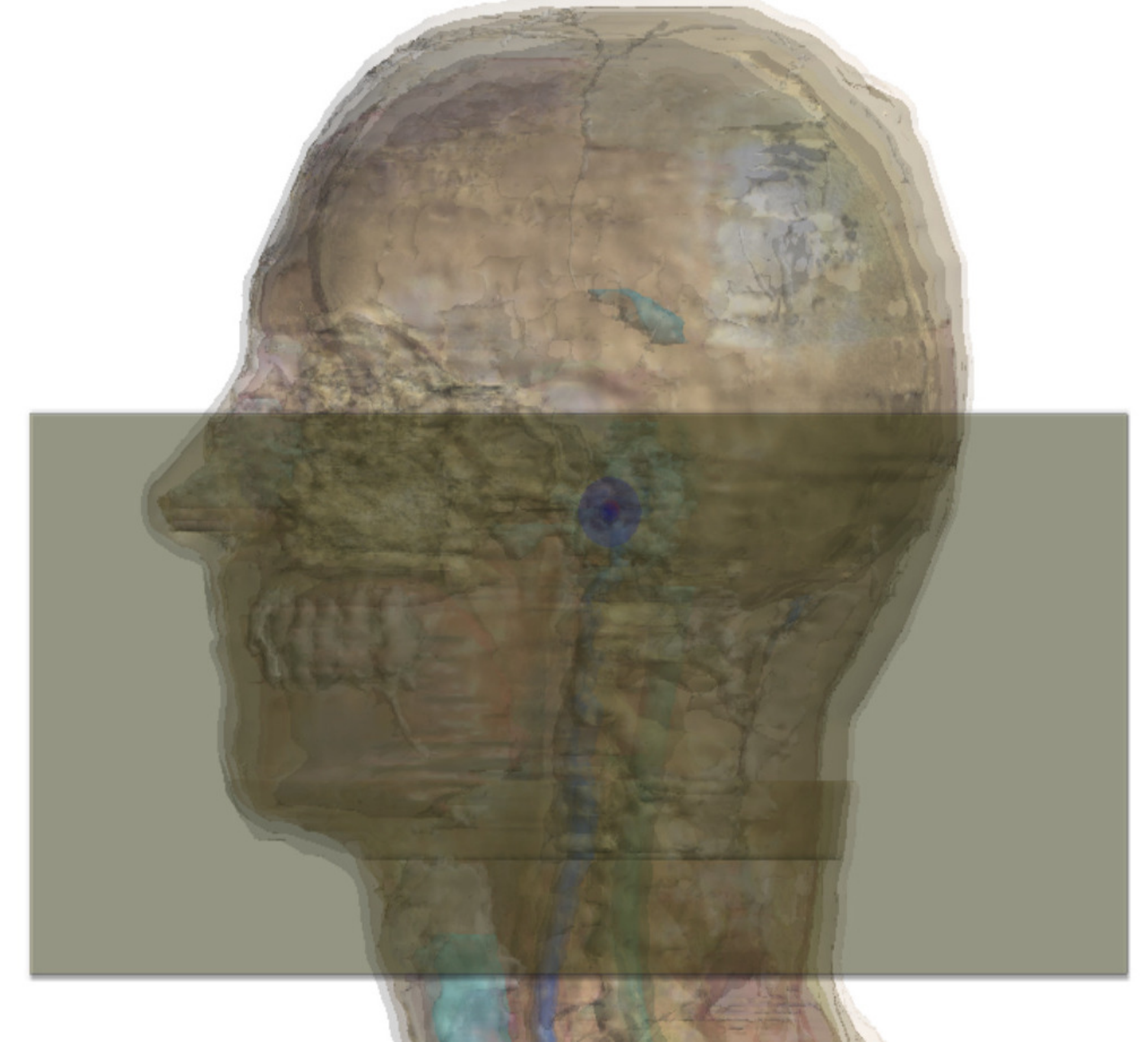}}
    			\subfigure[]{\label{case5}\includegraphics[scale=0.075]{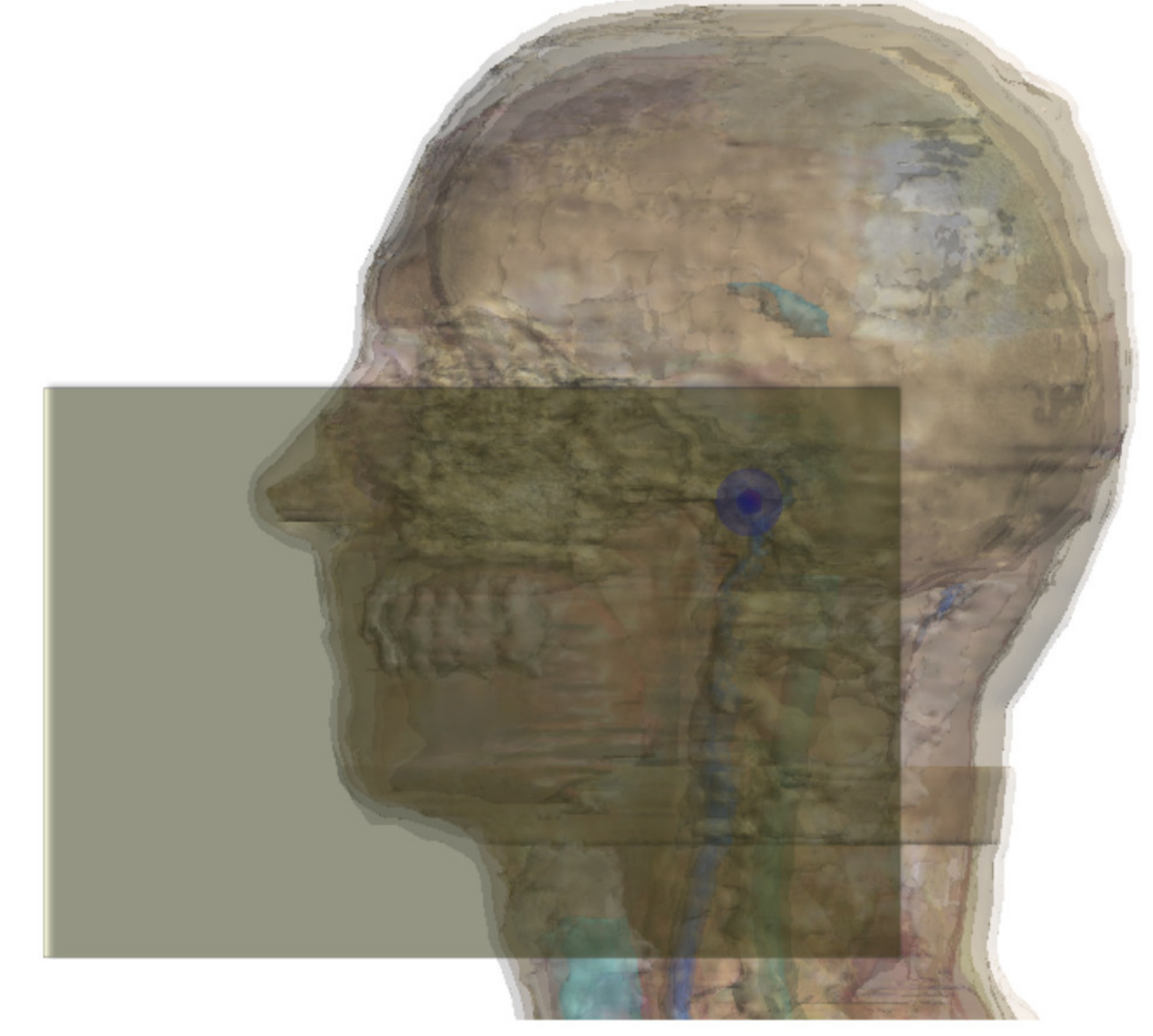}}
    			\subfigure[]{\label{case6}\includegraphics[scale=0.075]{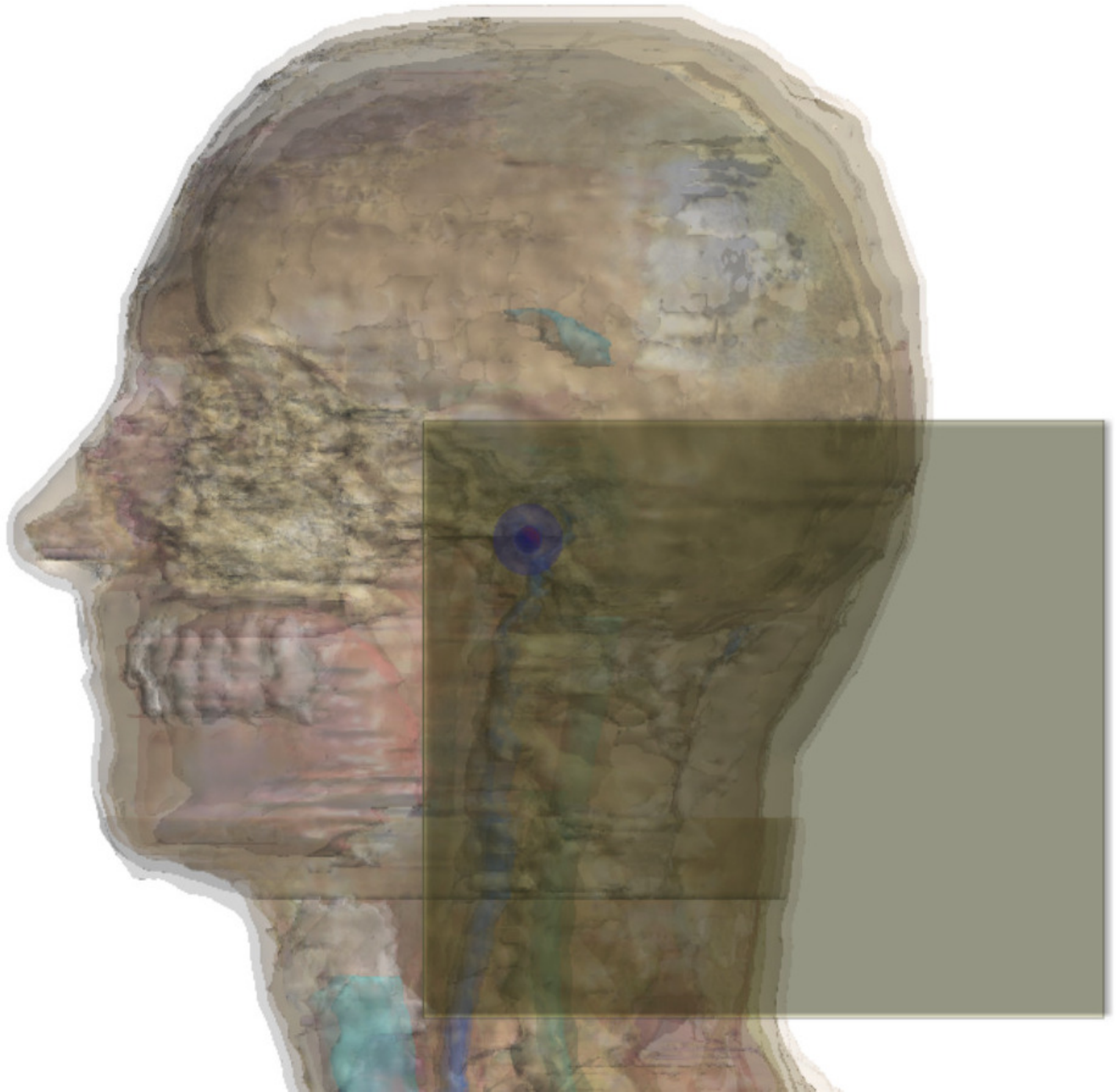}}

  \end{center}
      \caption{Duke head with the simulation domain for determining dominant creeping path. Shadowed region is the simulation domain. (a) Creeping path: front and top; Blocked: back (b) Creeping path: top; Blocked: front and back (c) Creeping path: back and top; Blocked: front (d) Creeping path: front and top; Blocked: back (e) Creeping path: front; Blocked: top and back (f) Creeping path: back; Blocked: front and top}%
     \label{PML_head}
   \end{figure}

\begin{figure}[thpb]
      \centering
      \includegraphics[scale=.20]{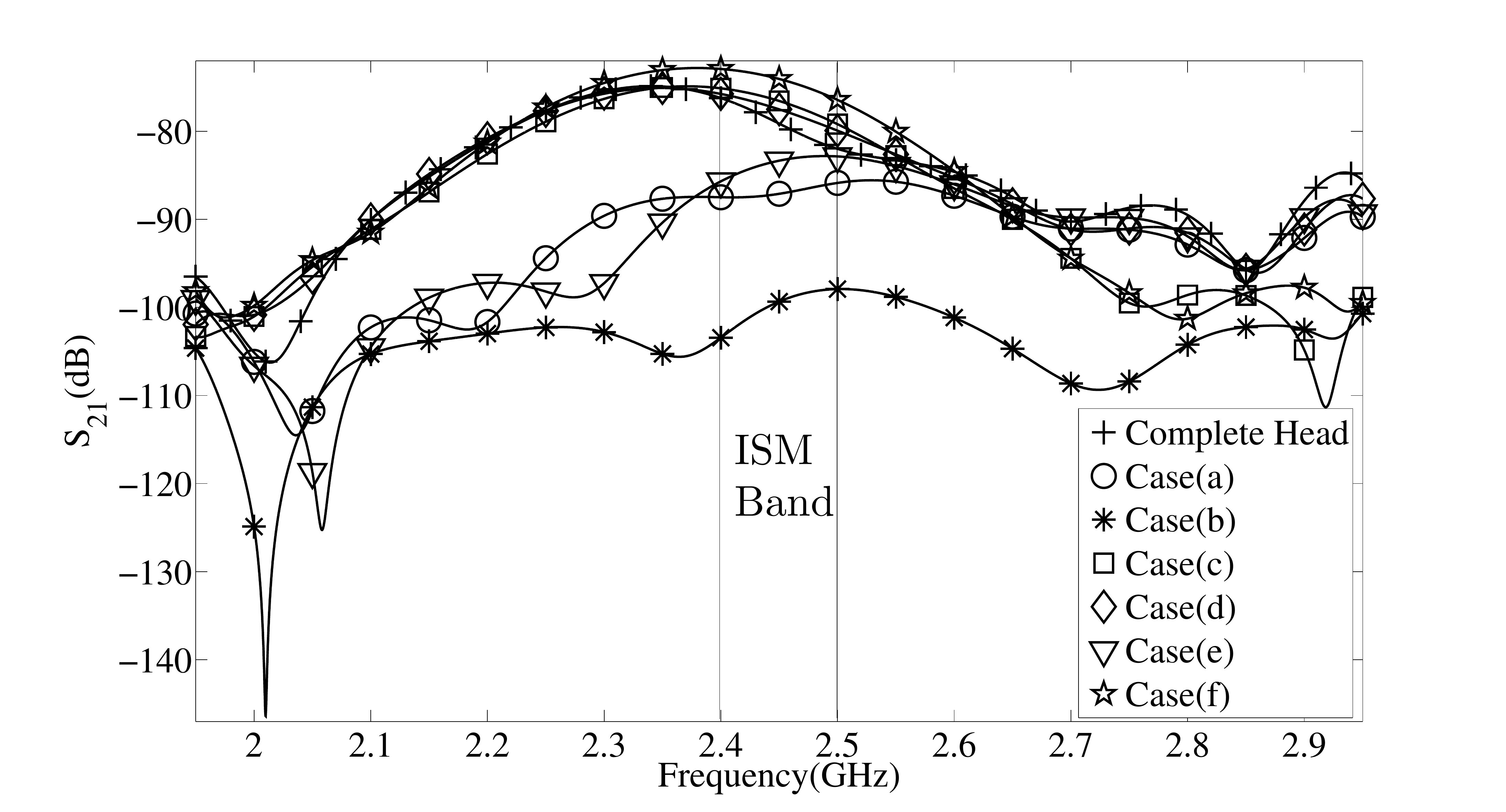}
      \caption{$S_{21}$ for different scenarios for determining the paths of the creeping waves carrying significant power. For the description of the cases in the legend refer Table~\ref{sim_cases}.}%
     \label{link_loss_pml}
   \end{figure}

The creeping waves from the transmitter in one ear can reach the receiver in the other ear by numerous paths. But it is not necessary that all these paths will have significant power as some paths will have high attenuation. Six different simulation cases are considered to determine the paths having significant power using the ITE antenna \cite{chandra1} on the Duke phantom. In these cases, one or more creeping paths are truncated by an uni-anisotropic perfectly matched layer (UPML) such that the creeping waves are blocked on these paths. The descriptions of the cases are presented in Table~\ref{sim_cases}. The table lists the creeping paths and the blocked paths for the six cases. For example, for case (a), simulation domain includes the front and the top part of the head, allowing the waves to creep over these paths and truncating the back side of the head with an UMPL which blocks the creeping waves on the back side of the head. It should be noted here that the front, the back and the top are relative position w.r.t. the ear (antenna placement). These six cases are illustrated in Fig.~\ref{PML_head} where the shadowed region represents the simulation domain. The link loss of all these cases is compared with the link loss when the complete head is used in the simulation domain. The simulation results are shown in Fig.~\ref{link_loss_pml}. Throughout the paper we present the plot of $S_{21}|_{dB}$ which is related to the link loss ($LL$) in dB scale by: $LL|_{dB} = -S_{21}|_{dB}$. It should be noted that the frequency band of interest is $2.45$ GHz ISM band and hence all the comparison are done in this band where the antennas are matched to $50~\Omega$ source impedance. At other frequencies like around $2$ GHz and $2.9$ GHz, there is an extreme dip in the simulated $S_{21}$ for some scenarios which is either because of the high mismatch loss at these frequencies or strong destructive interference of waves coming from different paths. However, since these are not the frequencies of the interest, no further investigations are done for these extreme dips in the simulated $S_{21}$.

\begin{table}[ht]\scriptsize
\caption{Simulation cases for determination of dominant path of creeping waves between the ears}
\label{sim_cases}
\begin{center}
\begin{tabular}{|c||c||c|}
\hline 
Scenario				  &		  Creeping Path	         &	 Blocked Path			                  \\                    
\hline
Case (a)           &     Front and Top        &   Back               					  \\
\hline
Case (b)						&			Top             	   &   Front and Back                 					\\
\hline
Case (c)					  &			Top and Back         &	  Front                 					\\
\hline
Case (d)						&			Front and Back       &    Top                					\\
\hline
Case (e)						&			Front           		 &		Top and Back                 	 		  	\\
\hline
Case (f)						&			Back                 &		Front and Top                					\\			
\hline
\end{tabular}
\end{center}
\end{table}    
   
\begin{table}[ht]\scriptsize
\caption{Difference between the link loss with and without UPML}
\label{diff_link_loss}
\begin{center}
\begin{tabular}{|c||c||c||c|}
\hline 
Scenario				    &		$\Delta S_{21}$(dB)	 &	 $\Delta S_{21}$(dB)		&		$\Delta S_{21}$(dB)	\\
                    &      [2.4 GHz]         &    [2.45 GHz]        		&      [2.5 GHz]        \\
\hline
Case (a)            &     11.3               &   8.0                    &   3.9 								\\
\hline
Case (b)						&			27.2             	 &   20.2                   &		15.9								\\
\hline
Case (c)					  &			-1.1            	 &	  -2.5                  &		-2.8								\\
\hline
Case (d)						&			-0.4            	 &    -1.6                  &		-2.1								\\
\hline
Case (e)						&			9.5            		 &		4.2                  	&		0.8 								\\
\hline
Case (f)						&			-3.3               &		-5.0                  &		-5.6								\\			
\hline
\end{tabular}
\end{center}
\end{table}    
   
The difference between the link loss with these six simulation cases and the complete head is shown in Table~\ref{diff_link_loss} at $2.4$ GHz, $2.45$ GHz and $2.5$ GHz where $\Delta S_{21} = S_{21}|_{complete\_head} - S_{21}|_{UPML}$. A positive $\Delta S_{21}$ means increase in the link loss and a negative $\Delta S_{21}$ means decrease in the link loss when compared to the actual case when the complete head is included. It can be seen that whenever the back of the head is excluded from the simulation boundary (cf. case (a), (b), (e)) so that no wave creeps over the back side of the head, the link loss increased by $4.2$-$20.3$ dB at the central frequency of $2.45$ GHz. From this, it can be concluded that the creeping wave propagating over the back of the head is the strongest. This can also be confirmed by case (f) as the link loss is lowest when only back path is included. However, when the top of the head is excluded from the simulation boundary as in case (d), the link loss decreased by $1.6$ dB at $2.45$ GHz. Hence, the top creeping wave does not carry any significant power. This observation is also supported by the increase in the link loss by $20.3$ dB at $2.45$ GHz in case (b) where the waves creeps only over the head top. In case (d), where the front and the back of the head is included, the difference is small. Hence, the front and the back path carry most of the significant power. With these observations we can conclude that the two paths, one going around the back and the other going around the front part of the head should give a good estimation of the ear-to-ear link loss.

\subsection{Effect of the outer lossy skin and the Pinna}
The presence of the outer lossy skin and the pinnas on the heterogeneous phantoms introduces extra loss when compared with the SAM head. In \cite{chandra2}, it was shown that for low profile antennas which are very close to the head surface, the loss because of the pinnas could be up to $13$ dB. The presence of the pinnas may introduce an additional loss as the strongest creeping wave which creeps over back part of the head passes through them and hence gets attenuated. The loss because of the pinnas could be theoretically calculated by the sum of the propagation loss in a lossy medium (pinna) and the reflection loss as described in \cite{chandra2}.

To verify the effect of the pinnas on the ear-to-ear link loss, an approximate pinna is modeled for the SAM phantom with the lossy shell having the electrical properties same as that of the SAM liquid which is approximately same for the ear-cartilage of the Duke phantom. The electrical properties of the pinna is assigned the same values (permittivity, $\varepsilon_r$ = $39.2$ and conductivity, $\sigma_e$ = $1.8$ S/m) as the SAM liquid to maintain the homogeneity. The height of the pinna is taken as $18$ mm which is same as that of the pinna in the Duke phantom. A semicircular structure of the pinna resulted in $7$ dB further increment in the link loss whereas a more realistic structure of the pinna resulted in $11$ dB increment in the link loss~\cite{chandra2}.

\begin{figure}[thpb]
      \centering
      \includegraphics[scale=.20]{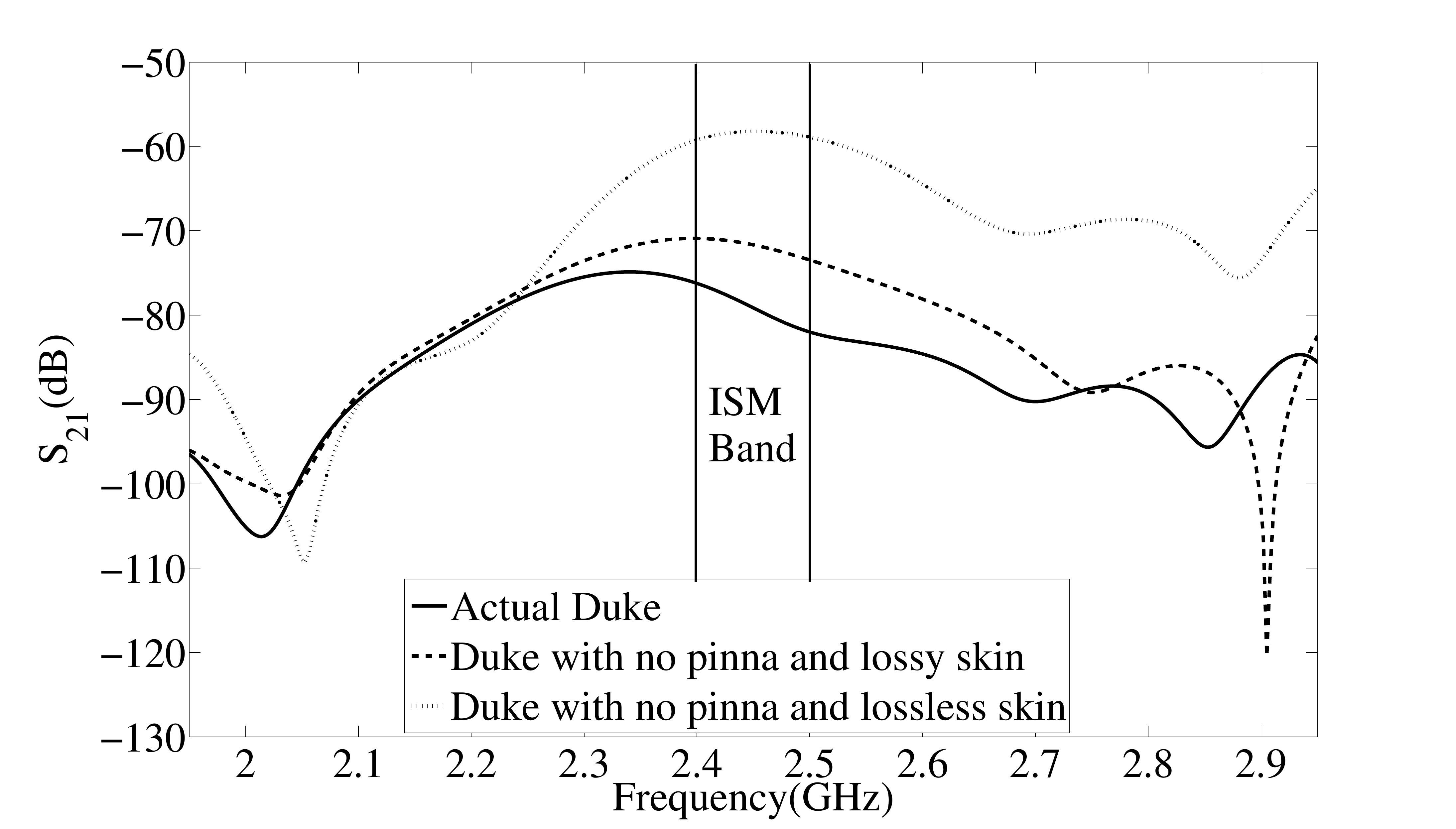}
      \caption{Effect of Pinna and Lossy Skin on Ear-to-Ear link loss for Duke}%
     \label{no_pinna_duke}
     \vspace{-1em}
   \end{figure}

A reverse methodology is implemented to verify the effect of the presence of the pinna and the lossy skin on the Duke phantom. The tissues of the ear in the Duke phantom, namely, the ear cartilage and the ear skin (see Table~\ref{tissue_parameter}) is assigned the electrical properties of air. This is done to remove the effect of the pinna. It resulted in decrease in the ear-to-ear link loss by $8$ dB at $2.45$ GHz. Next, the skin of the phantom (which is the outer most layer) is assigned the electrical properties of lossless SAM shell (permittivity, $\varepsilon_r$ = $3.7$ and conductivity, $\sigma_e$ = $0$). This further decreased the link loss by $13$ dB at $2.45$ GHz. The effects are illustrated in Fig.~\ref{no_pinna_duke}. 
 
\begin{figure}[thpb]
      \centering
      \includegraphics[scale=.20]{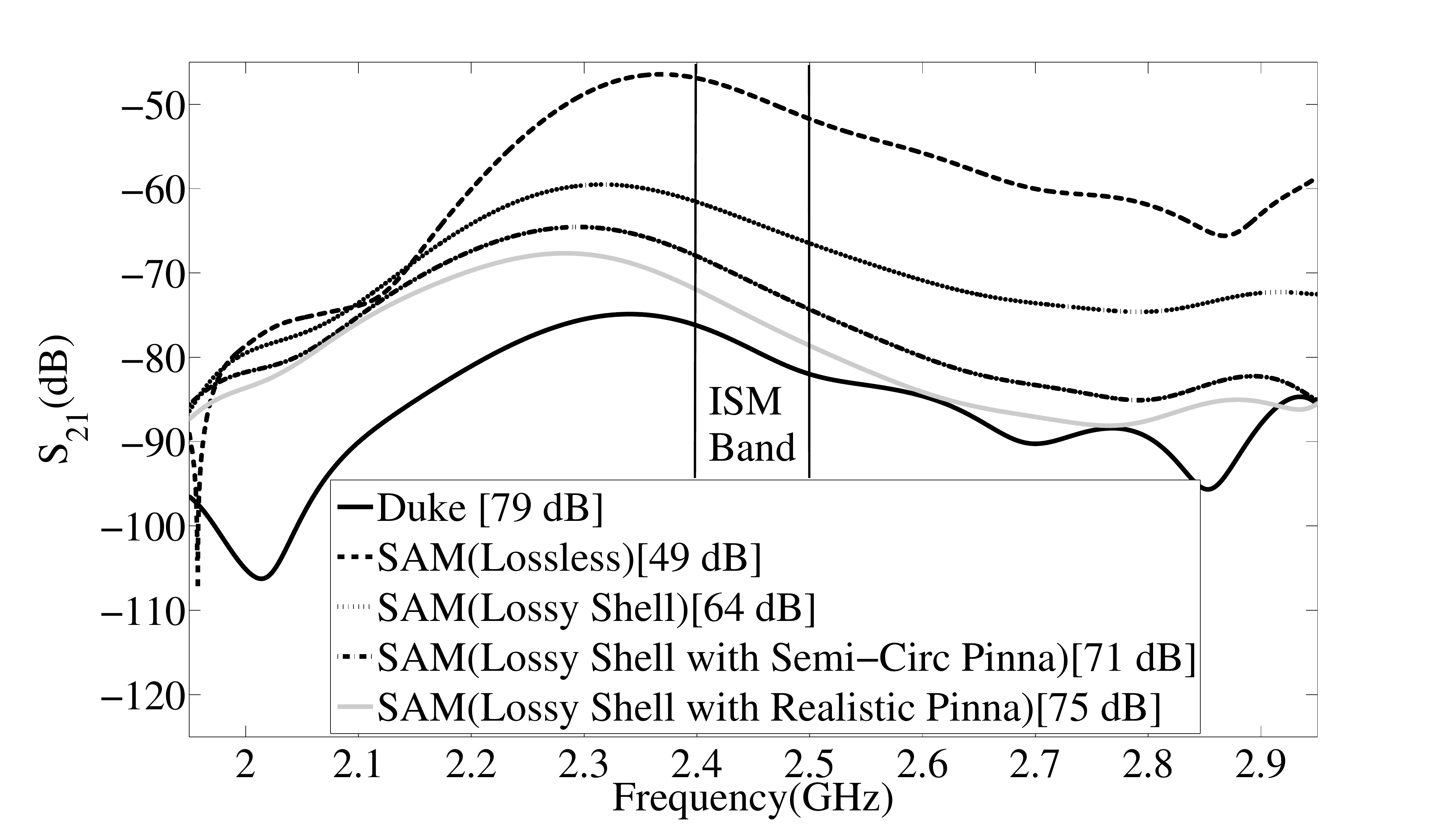}
      \caption{$S_{21}$ for Homogeneous vs. Heterogeneous Phantom using the ITE Antenna. The value in the square braces in the legend is the link loss at 2.45GHz.}%
     \label{link_loss_SAM_Duke}
     \vspace{-1em}
   \end{figure}

The link loss for the SAM phantom with a realistic pinna and a lossy shell approaches to that of the Duke phantom in the ISM band as seen in Fig.~\ref{link_loss_SAM_Duke}. Moreover, the reverse methodology shows the significant effect of the pinna and the lossy skin on the ear-to-ear link loss. Hence, it can be concluded that any numerical or real phantom with an outer lossless shell should be used with a caution while estimating the link loss for the on-body propagation and that the homogeneous phantoms should have a lossy outer shell. The phantoms should have the pinnas to get more accurate estimate of the ear-to-ear link loss. The shape of the pinna is also critical as more realistic pinna model estimates more accurate results.

   \begin{figure}[thpb]
      \centering
      \includegraphics[scale=.33]{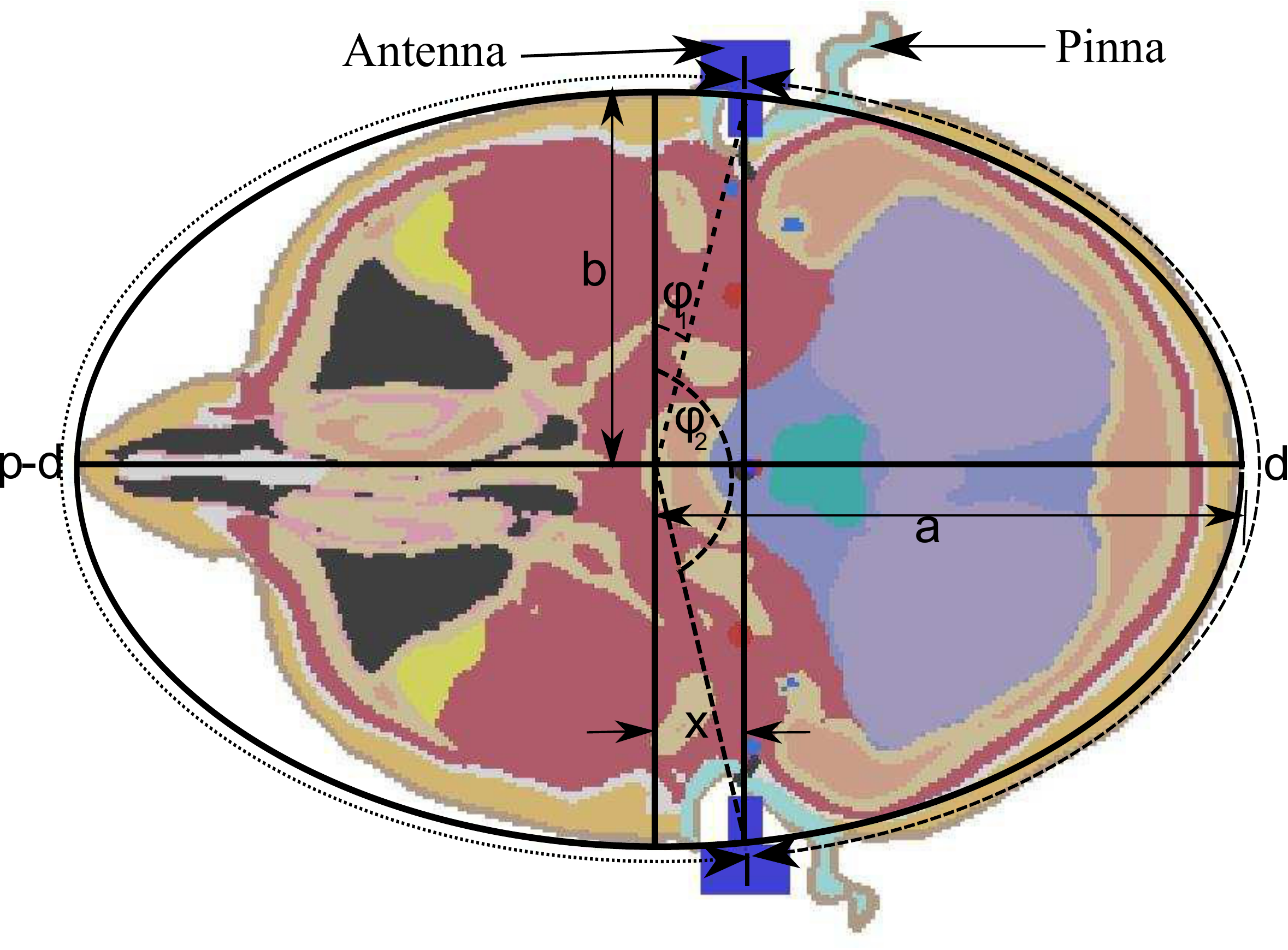}
      \caption{Elliptical fit for Duke in the transverse plane at the level of the antennas. $a = 115.8$ mm is the semi-major axis of the ellipse and $b = 75.1$ is the semi-minor axis of the ellipse. $d$ is the length of the path of the creeping wave going over the back side of the head and $p$ is the perimeter of the ellipse. $x$ is the offset of the ears from the center of the ellipse}%
     \label{duke_ellipse}
     \vspace{-1em}
   \end{figure}    
   
With the observations that the two paths of the creeping waves, namely, the front path and the back path are sufficient to describe the ear-to-ear link loss and the significant effect of the pinna and the lossy skin on the link loss, the analytical model for the ear-to-ear link loss is developed in the next section.
   
\subsection{Complete Link Loss Model}  
In this section the ear-to-ear link loss analytical model based on the creeping wave is developed. The model is for low profile antennas such that the height of the antenna above the body surface is negligible. The head is approximated by a more realistic elliptical cross-section rather than a circular cross-section. The elliptical approximation is illustrated in Fig.~\ref{duke_ellipse} for Duke head. In \cite{alomainy}, it was shown that the human body can be approximated by a metallic cylinder. Thus, the head is assumed to be metallic. The loss because of the waves creeping over the metallic surface is considered as an approximation for the waves creeping over the lossy medium. The lowest mode for the creeping wave is considered. A simpler version of the model for an elliptical approximation of the human torso was presented in our previous work \cite{chandraEucap2012}. Here, the model with the additional effect of the pinna is presented.

The creeping wave field on the elliptical path can be written as:
\begin{equation}\label{elecfield_ellipse}
{\bf E} = {\bf E}_0e^{-L}
\end{equation}
where the reference field ${\bf E}_0$ over the conducting surface for the vertical polarization at a distance $s$
is given by \cite{alves}:
\begin{equation}\label{eo}
{\bf E}_0 = 2\sqrt{\frac{\eta_0}{2\pi}}{\frac{\sqrt{P_{TX}G_{TX}}}{s}}e^{-jks}
\end{equation}
where $P_{TX}$ is the feeding power, $k=\frac{2\pi}{\lambda_0}$ is the wave number in the free space and $G_{TX}$ is the gain of the transmitting antenna.
$L$ is the complex attenuation factor representing the loss on the surface. It is product of creeping distance and creeping attenuation per unit length. For the elliptical path over a metallic surface, it is given by \cite{balanis}:
\begin{eqnarray}\label{l}
L = \frac{(k)^{1/3}}{2}\left(\frac{3\pi ab}{4} \right)^{2/3}e^{\frac{j\pi}{6}} &&\nonumber \\
.\int_{\varphi_1}^{\varphi_2}\frac{ab}{\sqrt{[a^4\cos^2\varphi + b^4\sin^2\varphi][a^2\cos^2\varphi + b^2\sin^2\varphi]}}\mathrm{d}\varphi
\end{eqnarray}
where $a$ is the semi-major axis and $b$ is the semi-minor axis of the ellipse. $\varphi_1$ and $\varphi_2$ are the exit point angle at the transmitter and the trapping point angle at the receiver respectively as shown in Fig.~\ref{duke_ellipse}. 

The total field at the receiver can be written as:
\begin{equation}\label{E_total}
{\bf E} = {\bf E}_{f} + {\bf E}_{b}
\end{equation}
where the subscript $f$ is for the wave creeping over the front part of the head and $b$ is for the back of the head. ${\bf E}_{f}$ is given by:
\begin{equation}\label{elecfield}
{\bf E}_{f} = {\bf E}_{0f}e^{-L_f}
\end{equation}
where the reference field ${\bf E}_{0f}$ for the front part of the head is given by (\ref{eo}) with $s = p - d$, 
where $p$ is the perimeter of the ellipse and $d$ is the elliptical arc length between the ears at back side of the head (as shown in Fig.~\ref{duke_ellipse}). The arc length, $t$, of the creeping wave between the the angle $\varphi_1$ and $\varphi_2$ on the elliptical surface can be expressed as \cite{balanis}:
\begin{equation}\label{arc_length}
t = ab\int_{\varphi_1}^{\varphi_2}\frac{(a^4\cos^2\varphi+ b^4\sin^2\varphi)^\frac{1}{2}}{(a^2\cos^2\varphi+ b^2\sin^2\varphi)^\frac{3}{2}}\mathrm{d}\varphi
\end{equation}

$L_f$ can be calculated by (\ref{l}) with the integration from $-\varphi_1$ to $\varphi_2 = \pi+\varphi_1$, where $\varphi_1$ is given by:
\begin{equation}\label{varphi1}
\varphi_1 = \tan^{-1}\left[\frac{a}{b}\frac{x}{\sqrt{a^2-x^2}}\right]
\end{equation}
where $x$ is the offset of the ear from the center of the ellipse as shown in Fig.~\ref{duke_ellipse}. $d$ can be calculated by substituting the proper values of $\varphi_1$ and $\varphi_2$ and $p$ can be calculated by substituting $\varphi_1 = 0$ and $\varphi_2 = 2\pi$ in (\ref{arc_length}). 
The wave going over the back of the head has to pass through the pinnas and hence there will be reflection and absorption loss apart from the surface attenuation. $E_{b}$ is given by:
\begin{equation}\label{elecfield_back}
{\bf E}_{b} = {\bf E}_{0b}e^{-L_b}T^2_{pinna}e^{-2\alpha R}
\end{equation}
The reference field, ${\bf E}_{0b}$ is given by (\ref{eo}) with $s = d$. 
For $L_b$ in (\ref{l}), $\varphi_1$ is same as that in (\ref{varphi1}) and $\varphi_2 = \pi-\varphi_1$. $T_{pinna}$ is the equivalent transmission coefficient for the air-pinna-air interface given by \cite{molisch} as:
\begin{equation}
\label{T}
T_{pinna} = \frac{T_1 T_2e^{-j\alpha_d}}{1+\rho_1 \rho_2 e^{-2j\alpha_d}}
\end{equation}
where $T_1$ and $T_2$ are the transmission coefficient of the air-to-pinna and pinna-to-air interface respectively. $\rho_1$ and $\rho_2$ are the reflection coefficient of the air-to-pinna and the pinna-to-air interface respectively. $\alpha_d$ is the electrical length of the pinna with permittivity $\epsilon_r$ for the normally incident waves and is given by $\alpha_d = \frac{2\pi\sqrt{\epsilon_r}R}{\lambda}$. Since the vertical polarization is dominant for the used antenna (described in the next section), TM waves are assumed. $e^{-\alpha R}$ in (\ref{elecfield_back}) represents the absorption loss in the pinna of average thickness $R$ with $\alpha = |Im[k_{pinna}]|$ where $k_{pinna}$ is complex wavenumber of the lossy medium (pinna). It should be noted that the terms $T_{pinna}$ and $e^{-\alpha R}$ are squared to represent the effect of the both pinnas.
The received power is:
\begin{equation}\label{E_total_e}
 P_{RX} = \frac{|{\bf E}|^2 A_{RX}}{2\eta_0}
\end{equation}

Substituting $A_{RX} = G_{RX}\frac{\lambda^2}{4\pi}$ and ${\bf E} = {\bf E}_{f} + {\bf E}_{b} $, the link loss $\left(\frac{P_{TX}}{P_{RX}} \right)$ between the perfectly matched antennas in the dB scale can be written as:
\begin{eqnarray}
\label{LinkLossdB}
LL|_{dB} &=& -10\mathrm{log}\left[\frac{G_{RX}G_{TX}\lambda^2}{4\pi^2}\left(\vline \frac{e^{-L_f}}{p-d}e^{-jk(p-d)}+\right.\right. \nonumber \\
&& \left.\left. \frac{e^{-L_b}T^2_{pinna}e^{-2\alpha R}}{d}e^{-jkd}\vline\right)^2\right]
\end{eqnarray}

\subsection{Verification of the Analytical Model}   
\begin{figure}[thpb]
      \centering
      \includegraphics[scale=.24, clip = true, trim=3cm 0cm 0cm 0cm]{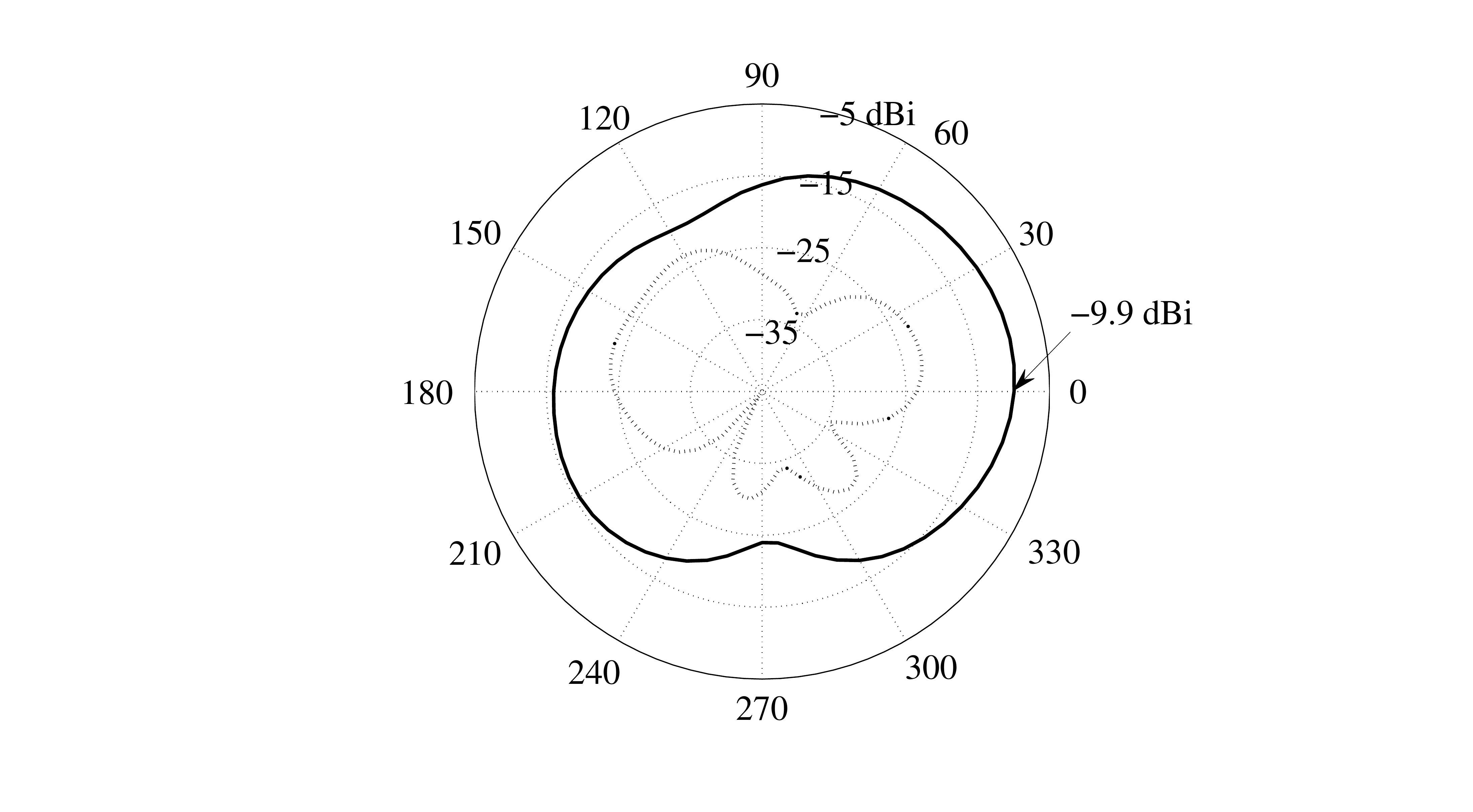}
      \caption{Gain Pattern of the antenna on Duke}%
     \label{Duke_Gain}
     \vspace{-1em}
   \end{figure}

\begin{figure}[thpb]
      \centering
      \includegraphics[scale=.21]{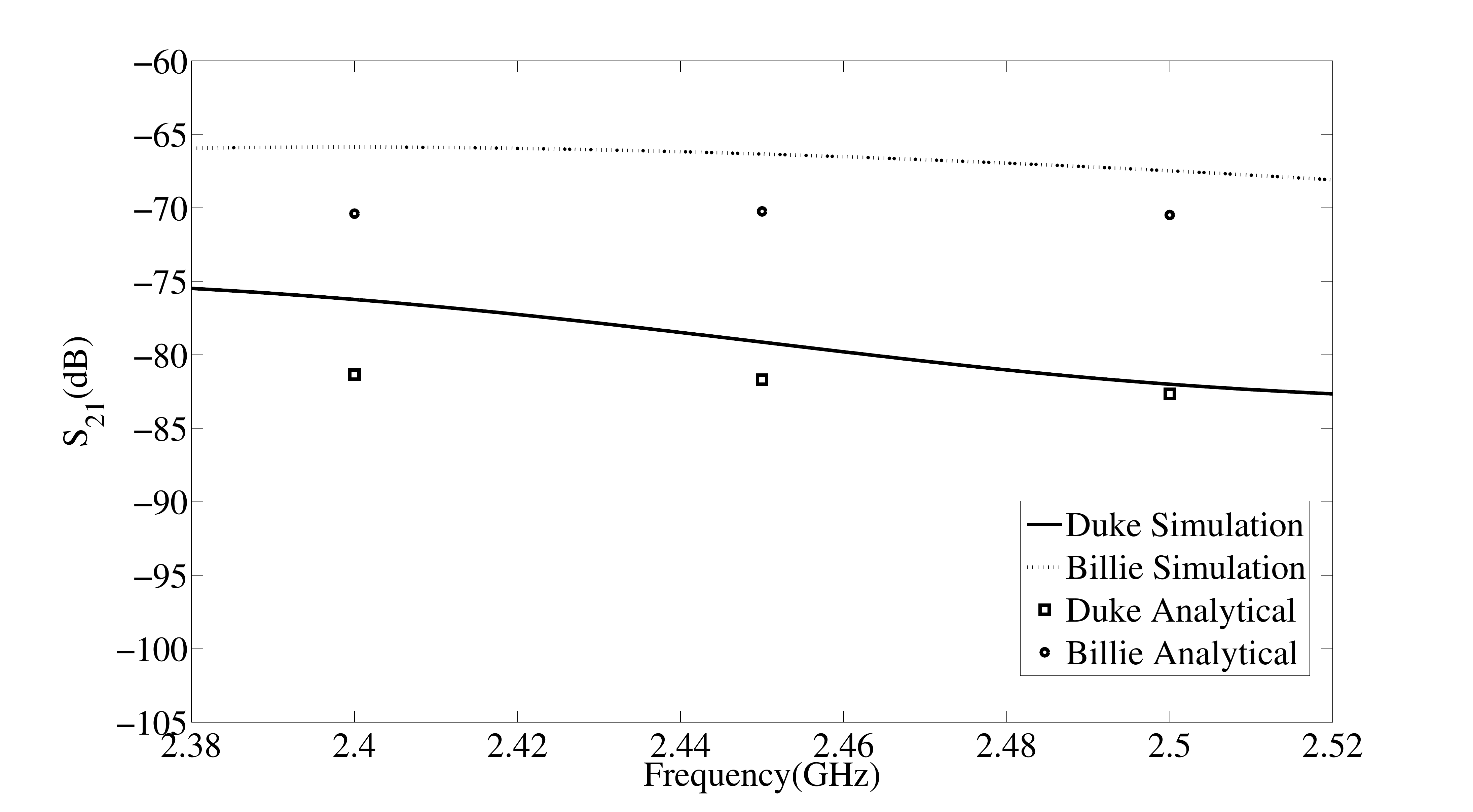}
      \caption{Link Loss: Simulated vs. Analytical}%
     \label{sim_ana}
     \vspace{-1em}
   \end{figure}

The analytical model is verified for the phantoms Duke and Billie, having the largest and the smallest head respectively. The parameters of Duke's and Billie's head after the elliptical fit are described in Table~\ref{analytical}. The values of $L_f$, $L_b$, creeping loss (losses excluding the gain of the antennas) and the the total ear-to-ear link loss obtained from the analytical model are also shown. The gain of the antenna is calculated from the simulations as described in \cite{alves}. The gain of the antenna in the azimuthal plane for the vertical polarized E-field (solid line) and horizontal polarized (dot line) is shown in Fig.~\ref{Duke_Gain}. It can be seen from the radiation pattern that the vertical polarization (normal to the head surface) is dominant. 
Also, the pattern is asymmetrical i.e. the two directions in which creeping waves exit or reach the antenna have different gain as one side has the pinna. Moreover, the two antennas on different ears have different gain due to the mismatch. Hence, the highest gain of the two antenna at $\varphi = 0^\circ$ is taken as the model assumes perfect match. The comparison between the simulated link loss and the link loss calculated with the analytical model in the ISM band is shown in Fig.~\ref{sim_ana}. The difference between the analytical and the simulated link loss is within $4$ dB for both the phantoms showing a good agreement.

     \begin{figure}[thpb]
      \centering
      \includegraphics[scale=.24, clip = true, trim=3cm 0cm 0cm 0cm]{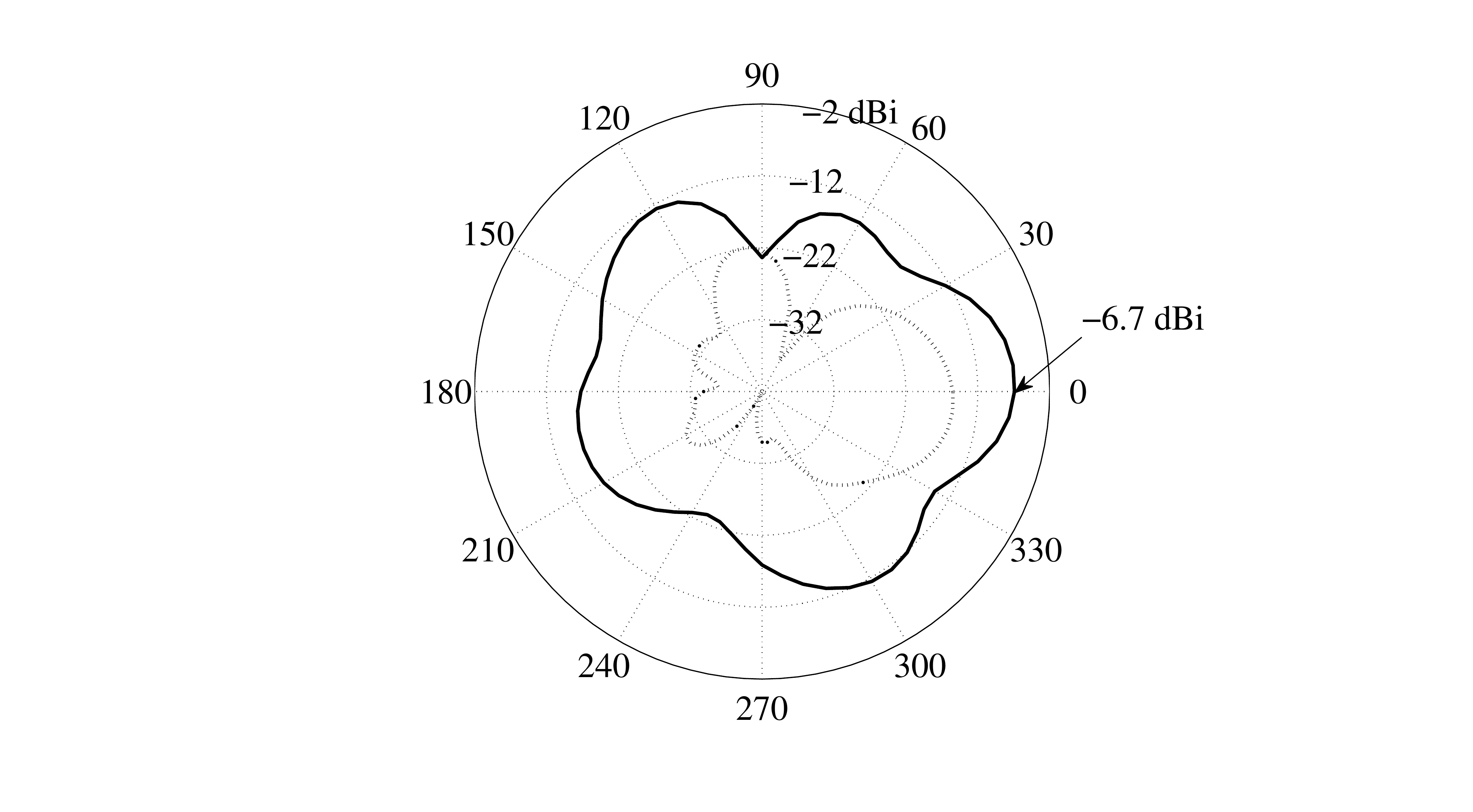}
      \caption{Gain Pattern of the antenna on Billie}%
     \label{Billie_Gain}
     \vspace{-1em}
   \end{figure} 
   
\begin{table}[ht]\scriptsize
\caption{Parameters of the Analytical Model}
\label{analytical}
\begin{center}
\begin{tabular}{|c||c||c||c||c||c|}
\hline 
Freq.(GHz)				    &		Gain(dBi)	 &	 $L_f$		&		$L_b$		&		CL$^{*}$(dB)		&		LL{$^\dagger$}(dB)\\
\hline
\multicolumn{6}{|c|}{{\bf{Duke}}: $a = 115.8$ mm, $b = 75.1$mm, $p = 609.7$ mm, $d = 270.8$ mm,}\\
\multicolumn{6}{|c|}{$x = 17$ mm, $R = 5$ mm, $\phi_1 = .225$ rad}\\
\hline
2.4			&			-10.0			&			4.30+2.48i			&			3.69+2.13i			&			61.34			&			81.34			\\
\hline
2.45		&			-9.95			&			4.33+2.50i			&			3.72+2.15i			&			61.80			&			81.70			\\
\hline
2.5			&			-10.2			&			4.36+2.52i			&			3.74+2.16i			&			62.27			&			82.67			\\
\hline
\multicolumn{6}{|c|}{{\bf{Billie}}: $a = 84.2$ mm, $b = 65.5$ mm, $p = 469.8$ mm, $d = 207.6$ mm,}\\
\multicolumn{6}{|c|}{$x = 13.7$ mm, $R = 4$ mm, $\phi_1 = .209$ rad}\\
\hline
2.4			&			-7.0			&			4.05+2.34i			&			3.37+1.95i			&			56.40		  &			70.40			\\
\hline
2.45		&			-6.7			&			4.08+2.35i			&			3.39+1.96i			&			56.85			&			70.25			\\
\hline
2.5			&			-6.6			&			4.10+2.37i			&			3.41+1.97i			&			57.29			&			70.49			\\
\hline
\end{tabular}
\end{center}
$^*$CL: Creeping Loss (losses excluding the antenna gain)\\
$^\dagger$LL: Total ear-to-ear link loss (including the antenna gain)
\end{table}

\subsection{Limitations and Discussions about the Analytical Model}
The analytical model gives the deterministic component of the ear-to-ear link loss. In the indoor scenarios where multi-path components (MPCs) are present, the link loss could be higher or lower depending upon the interference of the MPCs. The model is valid for the antennas which are very close to the human body so that the dominant propagation mechanism is through the creeping waves. The elliptical fit of the head is done when the models are looking parallel to the transversal plane. Duke was looking parallel to the transversal plane at the at the ear level (xy plane in the simulations) but for Billie, the model is rotated by $5^{\circ}$ to make her vision parallel to the xy plane.  The gain of the antenna is also critical as it is difficult to measure with the body \cite{alves}, but could result in significant differences in the link loss. The difference in the link loss from the analytical model between Duke and Billie is about $11$ dB at $2.45$ GHz. $5$ dB out of this $11$ dB is the difference in the creeping loss attributed to the difference in the head size and the ear thickness. The rest $6$ dB is due to the difference in the gain of the antennas. Hence, a more appropriate method of calculating the gain of the antennas with the body may result in more accurate results. Moreover, as discussed in \cite{chandraEucap2012}, there is no well defined measure (to best of our knowledge) such as the gain for the coupling from the antenna to the creeping wave over the human body which involves near-field effects. Hence, using the standard gain which is calculated in far-field is an effective approximation. 

\section{Other Factors Affecting the Ear-to-Ear Link Loss}\label{sec:factors}
\subsection{Effect of Different Head Sizes}
\begin{figure}[thpb]
      \centering
      \includegraphics[scale=.21]{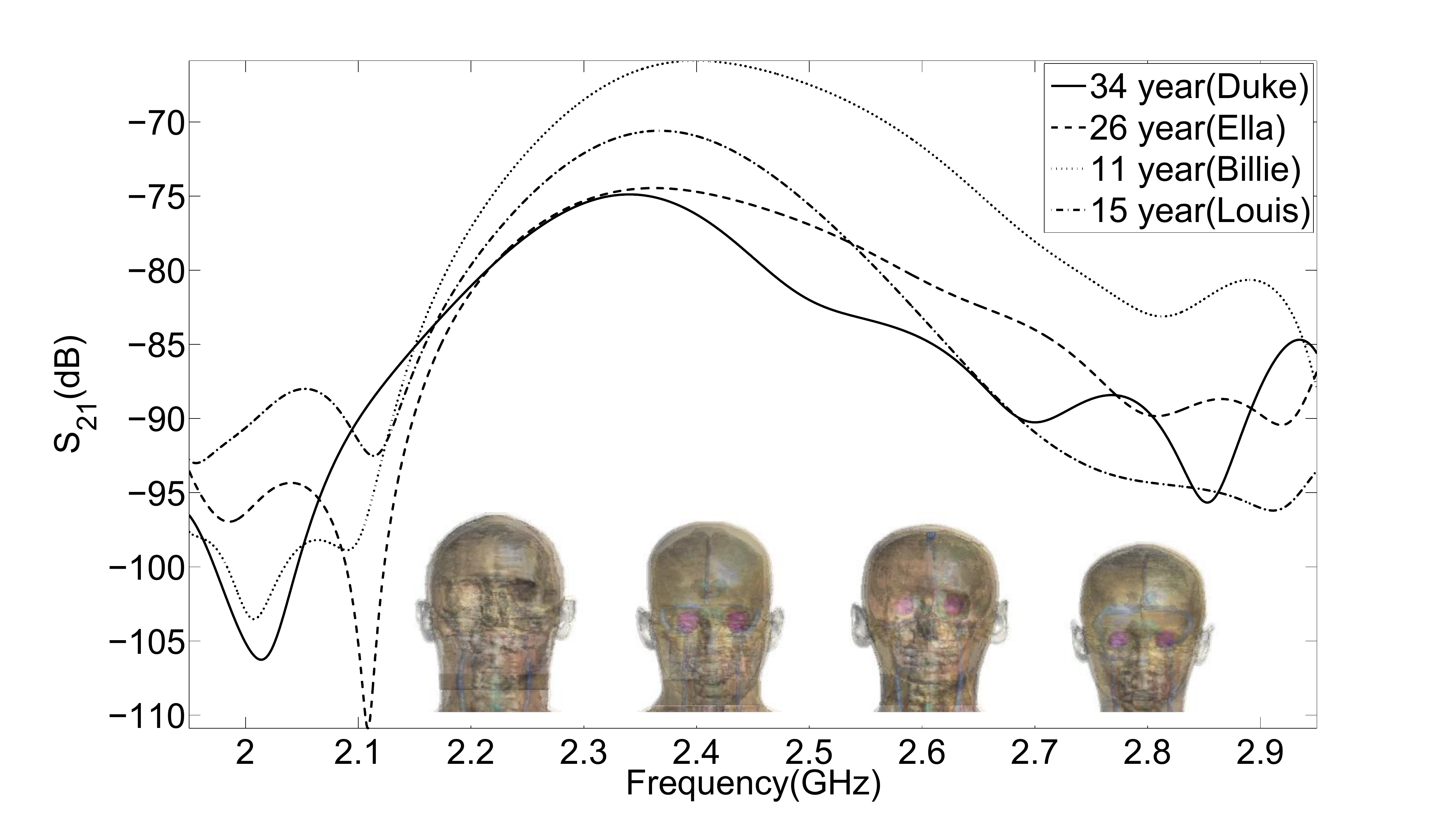}
      \caption{$S_{21}$ for different phantoms. From left to right: Duka, Ella, Louis, Billie}%
     \label{link_loss_phantoms}
     \vspace{-1em}
   \end{figure}

The effect of the head size on the ear-to-ear propagation channel has been discussed in \cite{kvist}, \cite{kvist1} and \cite{chandra2}. In our previous work \cite{chandra2}, we have shown by the simulations on the heterogeneous phantoms that different head size will have different link loss as shown in Fig.~\ref{link_loss_phantoms}. This can be attributed to the fact that the different head size will result in different path lengths of the creeping waves going behind the head and front of the head and also because of the gain variations of the antenna resulting from anatomical variations in different phantoms. 

\subsection{Effect of the Shoulders} 
\begin{figure}[thpb]
      \centering
      \includegraphics[scale=.19]{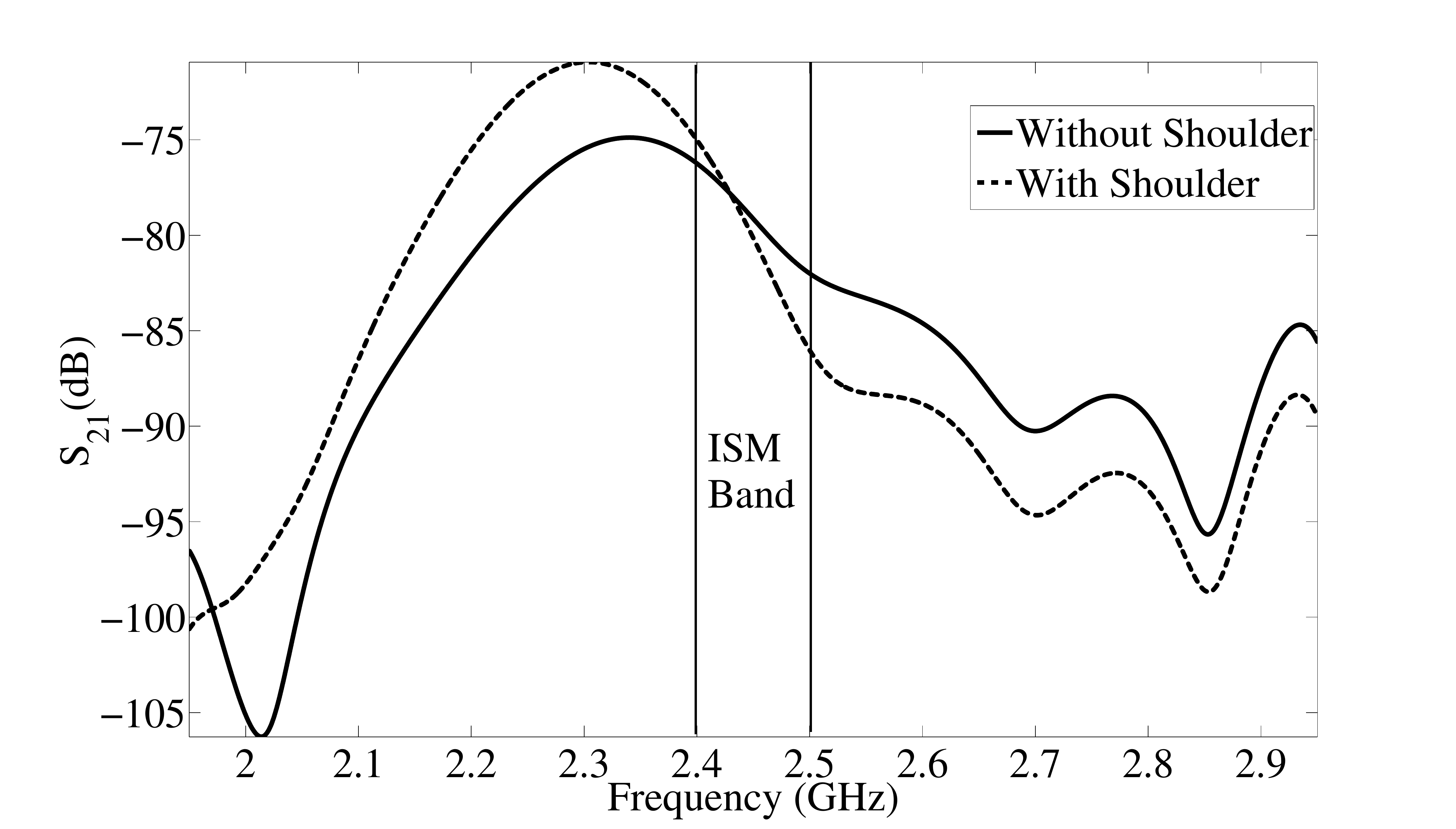}
      \caption{$S_{21}$ with shoulders and without shoulders for Duke phantom}%
     \label{shoulder1}
   \end{figure}

The shoulders may also affect the ear-to-ear link loss as the reflections from the shoulders may result in multipath components interfering either constructively or destructively with the creeping waves resulting in different power levels. The effect of the shoulders on Duke phantom can be observed in Fig.~\ref{shoulder1}. It can be seen that the difference between the link loss with and without the shoulders is within $2$ dB in the ISM band. Thus, there is a possibility to exclude the shoulders in order to decrease the simulation time~\cite{chandra2}. 

%
%
\section{Measurements}\label{sec:meas}
Measurements are carried out to verify the effect of the pinna. The measurement setup and the measured scenarios are described below. 
\subsection{Measurement Setup}
\begin{figure}[thpb]
      \centering
      \includegraphics[scale=.10]{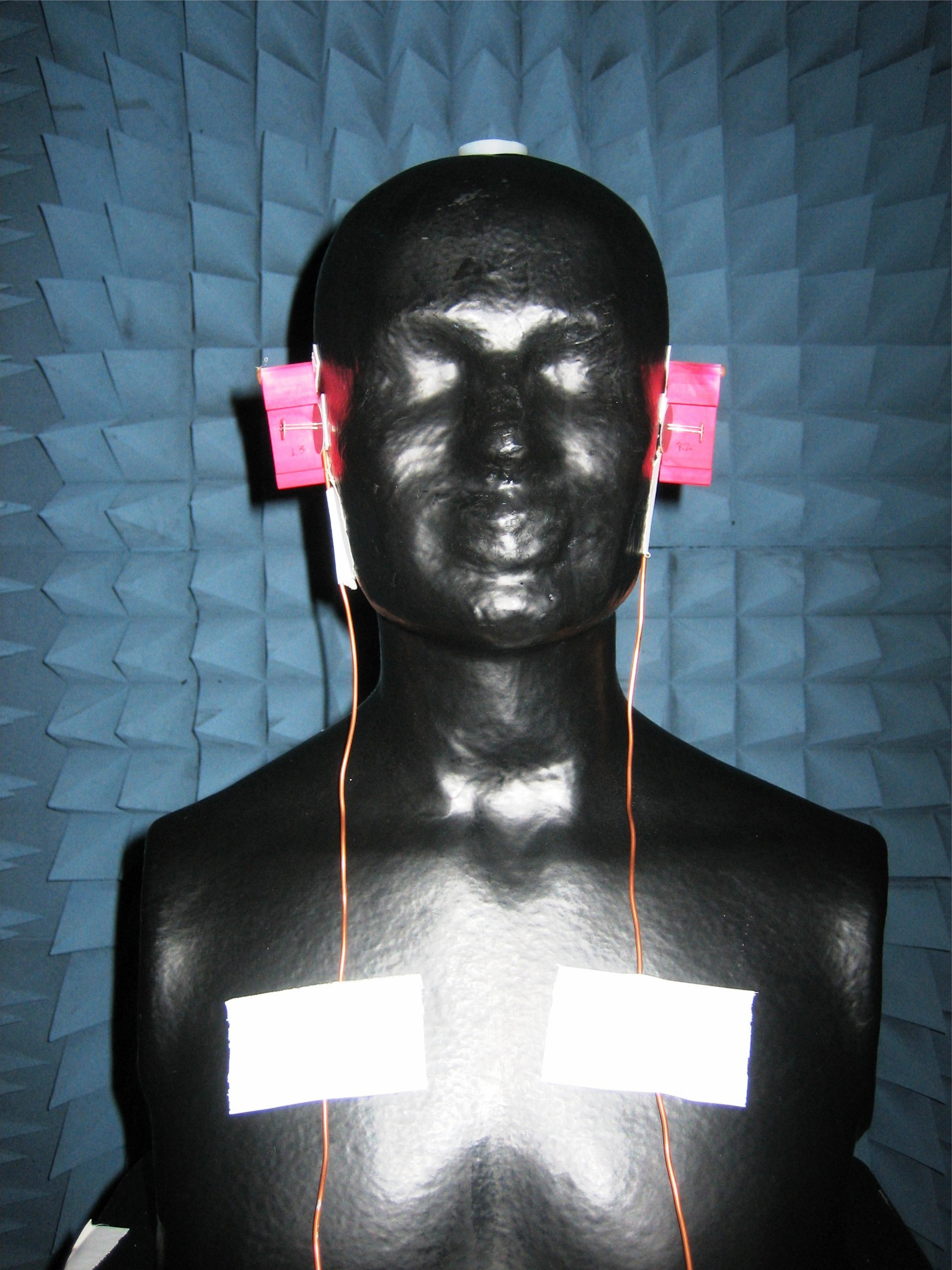}
      \caption{Phantom used for the measurement. The placement of the antennas and the pinna is can also be seen.}%
     \label{phantom}
   \end{figure}

The phantom shown in Fig.~\ref{phantom} is used for the measurements. The phantom is hollow and is filled with the tissue stimulating liquid mimicking the average electrical properties of the human head tissues at $2.45$ GHz (permittivity, $\varepsilon_r$ = $39.2$ and conductivity, $\sigma_e$ = $1.80$ S/m). The phantom does not have the pinnas. External pinnas are manufactured and attached to the phantom. The measured scenarios are (a) ear-to-ear link loss without the pinnas (b) ear-to-ear link loss with the pinna. Return loss of the antennas are also measured in all these scenarios. The measurement is done in an anechoic chamber of the EIT department, Lund University. S-parameters, $S_{11}$ and $S_{21}$, are measured at $1601$ frequency points in $1.95$ GHz to $2.95$ GHz frequency band using HP-8720C Vector Network Analyzer (VNA). The output power is set to $-10$ dBm. TOSM calibration is done to calibrate the VNA using a fabricated calibration kit. The calibration kit is fabricated in order to remove the effect of the semi-rigid cable which is attached to the antenna. 

\subsection{Fabricated Pinna}
\begin{figure}[thpb]
      \centering
      \includegraphics[scale=.09]{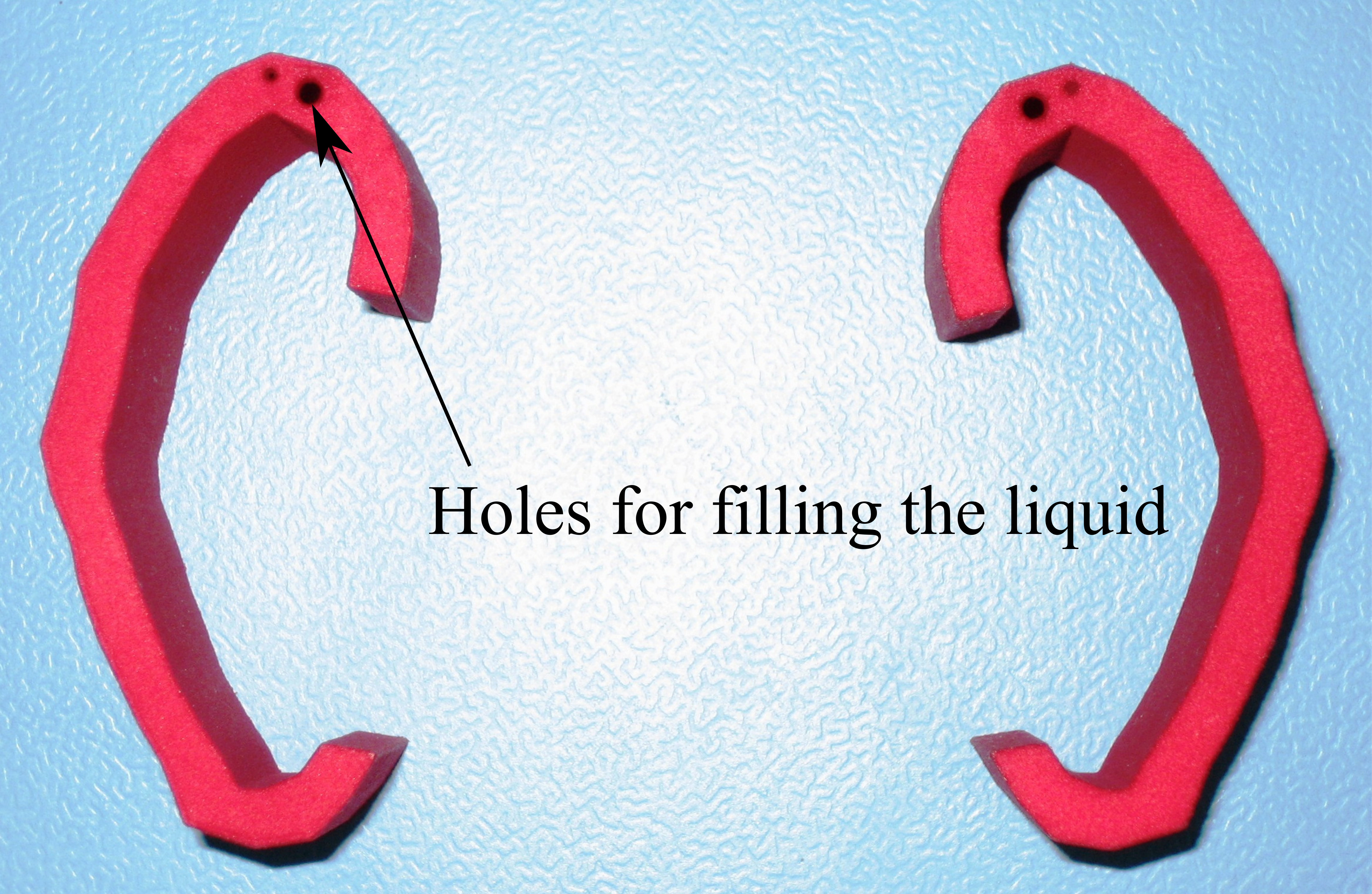}
      \caption{Pinnas fabricated by the 3D-printing}%
     \label{pinna_fab}
   \end{figure}

The pinnas are commercially fabricated by the 3D-printing technology by Shapeways \cite{shapeways}. Plastic is used as the material. They are hollow from inside and holes are made on them to facilitate the filling of the hollow pinnas with the tissue stimulating liquid. After filling the pinnas with the liquid, the holes are closed with a glue gun. The pinnas are glued on the phantom with the help of a thin double sided tape. The fabricated pinnas are shown in Fig~\ref{pinna_fab}. The phantom with the pinna and the antenna placement is shown in Fig~\ref{phantom}.

\subsection{Antennas}
\begin{figure}[thpb]
      \centering
      \includegraphics[scale=.20]{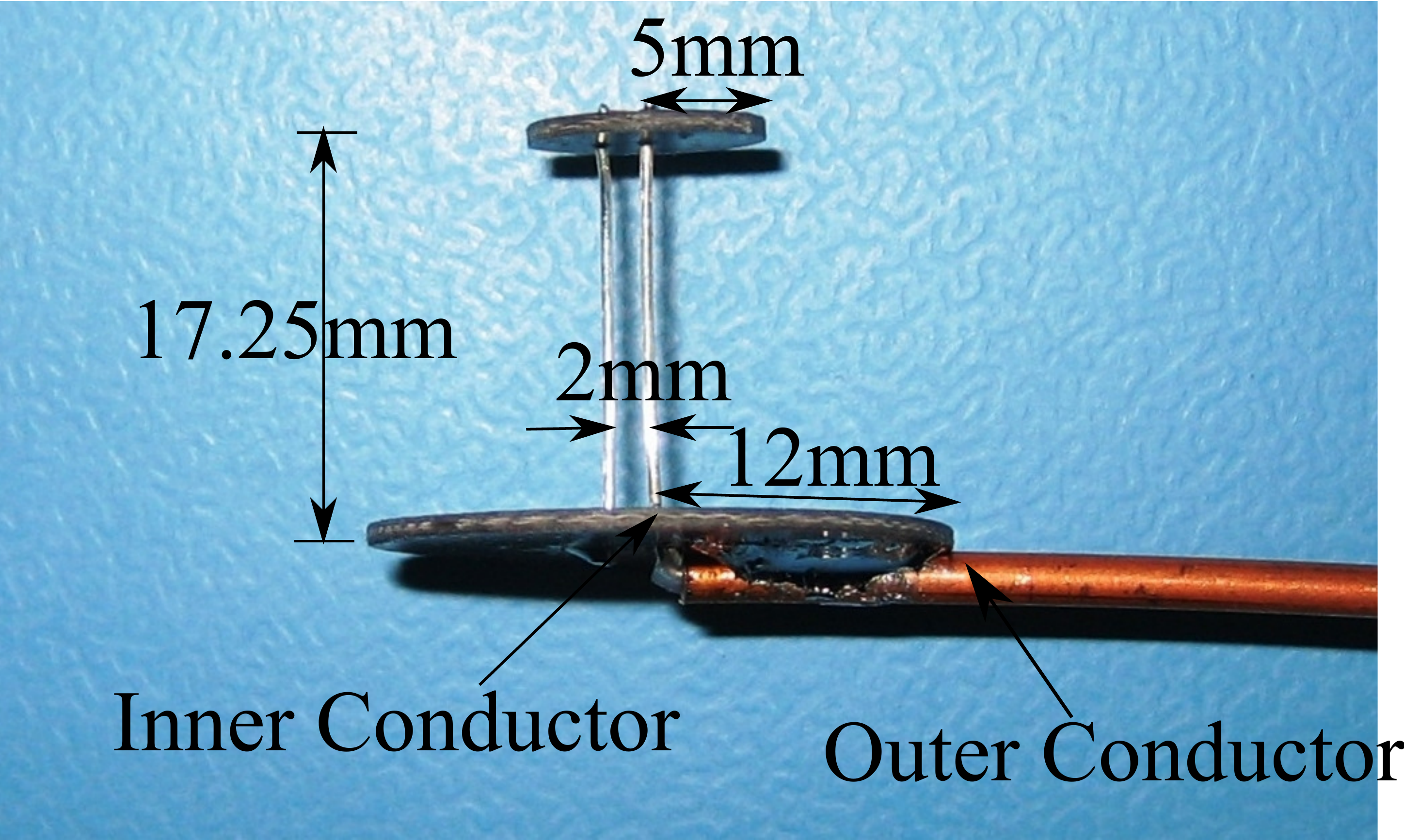}
      \caption{Fabricated Antenna}%
     \label{fab_antenna}
   \end{figure}

A monopole antenna is designed and optimized on the SAM phantom with the pinnas to work in $2.45$ GHz band. The antenna is loaded with a disc and then the disc is shorted to the ground plane to miniaturize the size and make the antenna low profile. The fabricated antenna with the dimensions is shown in Fig.~\ref{fab_antenna}. The central conductor of the semi-rigid coaxial cable, UT-85-H-M17 provided by Rosenberger \cite{rosenberger} is used as the central conductor in the monopole antenna. The outer conductor is then soldered to the ground plane of the antenna (also shown in Fig.~\ref{fab_antenna}). The length of the cable is $1$ m with the SMA connector at the other end of the cable. Hence, the calibration kit is fabricated to remove the affect of this semi-rigid cable. The semi-rigid cable has outer diameter of $2.2$ mm. It is used in order to place the antenna close to the head because directly attaching the SMA connector to the antenna could have resulted in much more air gap between the antenna and the head. The ground plane and the loading disc are etched out from a Teflon laminate of $0.8$ mm thickness having $0.35$ $\mu$m thick copper layer.
   


\subsection{Measured Results and Discussions} 
\begin{figure}[thpb]
      \centering
      \includegraphics[scale=.20]{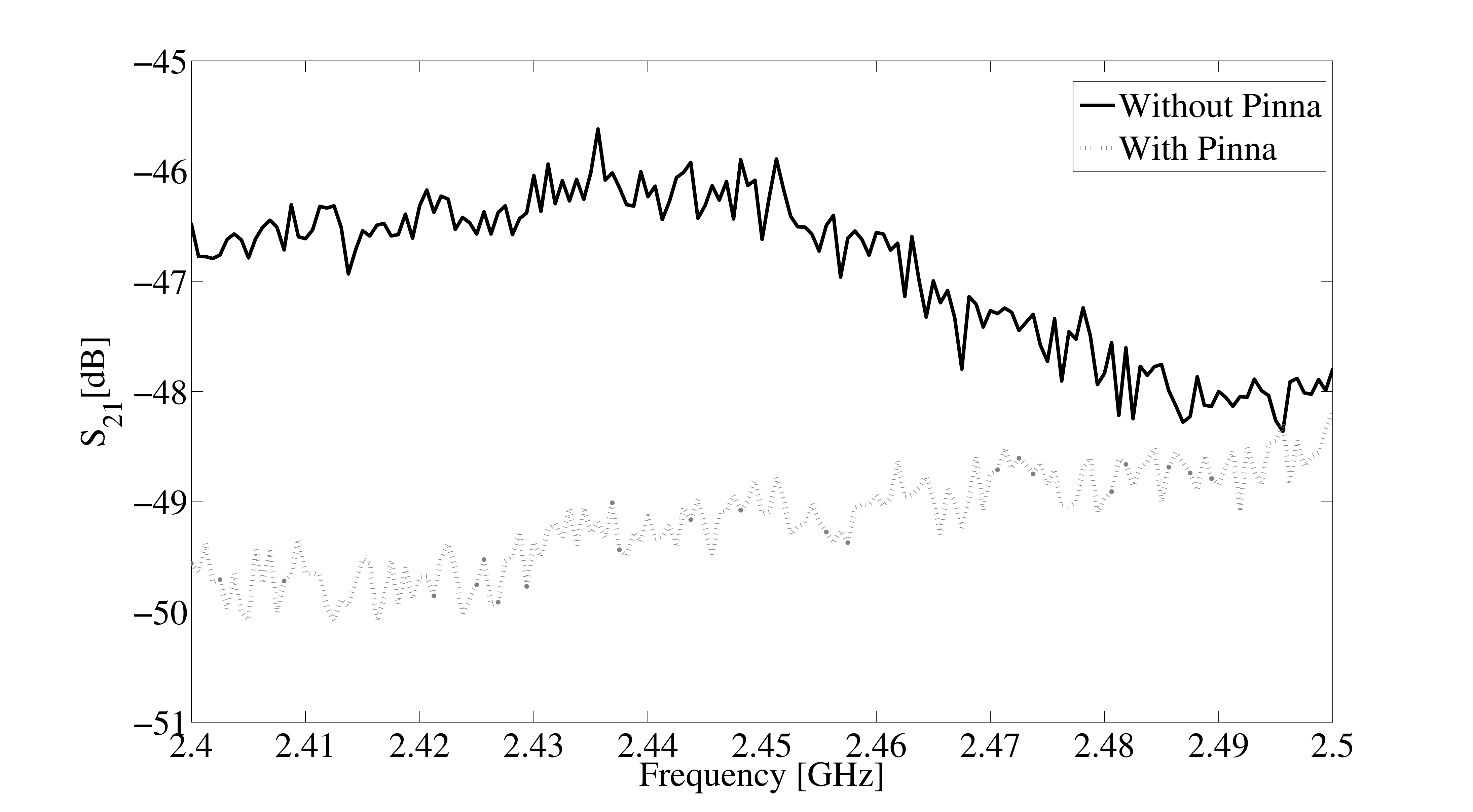}
      \caption{Measured $S_{21}$ with and without the pinna in the $2.45$ GHz ISM Band}%
     \label{meas_s21}
   \end{figure}

The return loss of the fabricated antenna is measured on the phantom, with and without the pinnas. 
It is observed that the antenna is mismatched when placed on the phantom without the pinna but the presence of the pinna improves the matching. The mismatch loss in the absence of the pinnas are removed from the measured $S_{21}$ for proper comparison by treating the return loss in the presence of the pinnas as the reference. This is done as:

\begin{equation}\label{s21_comp}
|S_{21}|_{no\_pinna}|^{2} = |S_{21}|_{no\_pinna}|_{meas}^{2} \times F
\end{equation}
where $F$ is the correction factor for removal of the mismatch loss without the pinna with the antenna with the pinna as the reference, given by:
\begin{equation}\label{mismatch_norm}
F = \frac{(1-|S_{11}|_{pinna}^{2})\times (1-|S_{22}|_{pinna}^{2})}{(1-|S_{11}|_{no\_pinna}^{2}) \times (1-|S_{22}|_{no\_pinna}^{2})}
\end{equation}

where $|.|$ represents the absolute value of the S-parameters in linear scale. The measured results are shown in Fig.~\ref{meas_s21}. It can be seen that the link loss without the pinna is $4$ dB less than the link loss with the pinna at $2.45$ GHz. The difference is not as significant as discussed in Section III(B). The possible reason for this could be the fact that the antennas used for the measurements are not low profile as the antennas used in simulations. Moreover, the phantom used for the measurement does not have the ear canals. Hence, the antennas are not shadowed to the extent as that in simulations. Nevertheless, the increase in the link loss because of the pinnas can still be observed.  


\section{Conclusions}\label{sec:conclusion}
An analytical model based on creeping wave for the ear-to-ear link loss was presented. It is useful in estimating the deterministic component of the link loss between the transceivers for binaural hearing aids. The model takes into account the losses on the surface of the head, the loss due to the pinna and models the cross-section of the head with an elliptical shape. Through the simulations it was observed that the two paths of the creeping wave, one going over the back of the head and the other in front of the head at the level of the ears were the dominant paths of the creeping waves carrying most of the power. Hence, a two path analytical model was developed. The verification of the model was done through the simulations on two heterogeneous phantoms of different age (34 year and 11 year) and head size. The difference between the model and the simulations was found to be within $4$ dB in the $2.45$ GHz ISM band showing a good agreement. The main benefit of the model lies in fast estimation of the link loss when compared with time and memory consuming numerical simulations. With the estimated value of the link loss, the sensitivity of the hearing aids could be decided. A comparison between the SAM phantom and a heterogeneous phantom was done which showed a significant effect of the pinna and the lossy skin. It was observed that the presence of the pinnas increased the link loss. This effect of the pinnas was verified through the measurements on a phantom where external pinnas manufactured by 3D-printing were attached. 



%

%
%
%
\addtolength{\textheight}{ 0cm} 

\section*{Acknowledgment}

This research work has been done as a part of the Ultra
Portable Devices (UPD) project. The authors would like to
thank Swedish Foundation for Strategic Research (SSF) for
funding the Ultra Portable Devices (UPD) project at Lund
University

\ifCLASSOPTIONcaptionsoff
  \newpage
\fi



%
%
\vspace{-2cm}
\end{document}